\documentclass[reprint,aps,superscriptaddress,amsmath,amssymb,article]{revtex4-2}
\usepackage{bm}        % for math
\usepackage{amssymb}   % for math
\usepackage{color}
\usepackage{units}
\usepackage{amsmath}
\usepackage{natbib}
\usepackage{esint}
\usepackage{chemformula}
\usepackage{ulem}
\usepackage[english]{babel}
\usepackage[colorlinks=true, citecolor=blue]{hyperref}
\usepackage{xcolor}
\DeclareUnicodeCharacter{2009}{\,}
\DeclareUnicodeCharacter{2032}{\,}
\usepackage{chemformula} % Formula subscripts using \ch{}
\usepackage[T1]{fontenc} % Use modern font encodings
\usepackage{tikz}
\usepackage{parskip}
\usepackage{multirow}
\usepackage{mathtools} 
\usepackage{amsmath}

\usepackage{xcolor,colortbl}

\definecolor{Gray}{gray}{0.85}
\definecolor{LightCyan}{rgb}{0.88,1,1}

\newcolumntype{a}{>{\columncolor{Gray}}c}
\newcolumntype{b}{>{\columncolor{white}}c}

  \newcommand{\IISc}{\affiliation{
		 Department of Physics, Indian Institute of Science, Bangalore 560012, India}}
  
\newcommand{\IACS}{\affiliation{School of Physical Sciences, Indian Association for the Cultivation of Science, Jadavpur, Kolkata 700032, India}}

\begin{document}
\title{Raman signatures of lattice dynamics across inversion symmetry breaking phase transition in quasi-1D compound, \ch{(TaSe4)3I} }
%\title{ Raman signatures of inversion symmetry breaking in quasi-1D compound, \ch{(TaSe4)3I} }
%\title{Raman analysis of strain-coupled lattice dynamics induced inversion symmetry breaking phase transition in quasi-1D compound, \ch{(TaSe4)3I} }
%\title{Evolution of collective lattice dynamics across inversion symmetry breaking transition in a quasi-1D compound (TaSe$_4$)$_3$I}
%\title{Probing inversion symmetry breaking transition in a quasi-1D compound (TaSe$_4$)$_3$I by investigating Raman signatures of collective lattice dynamics }
%\title{Broken inversion symmetry in a quasi-1D compound (TaSe$_4$)$_3$I, investigated by Raman spectroscopy. }
%% 	\title{Tracking Structural Phase Transitions in \ch{(TaSe4)3I} by Means of Phonon Mode Modification}
\author{Arnab Bera}
\altaffiliation{These authors contributed equally to this work}
\IACS

\author{Partha Sarathi Rana}
\altaffiliation{These authors contributed equally to this work}
\IISc

\author{Suman Kalyan Pradhan}
\IACS

\author{Mainak Palit}
\IACS
\author{Sk Kalimuddin}
\IACS

\author{Satyabrata Bera}
\IACS

\author{Tuhin Debnath}
\IACS

\author{Soham Das}
\IACS

\author{Deep Singha Roy}
\IACS

\author{Hasan Afzal}
\IACS

\author{Subhadeep Datta}
\IACS

\author{Mintu Mondal}
\email{mintumondal.iacs@gmail.com}
\IACS

\begin{abstract}
Structural phase transition can occur due to complex mechanisms other than simple dynamical instability, especially when the parent and daughter structure is of low dimension. This article reports such an inversion symmetry-breaking structural phase transition in a quasi-1D compound (TaSe$_4$)$_3$I  at T$_S\sim$ 141~K studied by Raman spectroscopy. Our investigation of collective lattice dynamics reveals three additional Raman active modes in the low-temperature non-centrosymmetric structure. Two vibrational modes become Raman active due to the absence of an inversion center, while the third mode is a soft phonon mode resulting from the vibration of Ta atoms along the \{-Ta-Ta-\} chains. Furthermore, the most intense Raman mode display Fano-shaped asymmetry, inferred as the signature of strong electron-phonon coupling. The group theory and symmetry analysis of Raman spectra confirm the displacive-first-order nature of the structural transition.  Therefore, our results establish (TaSe$_4)_3$I as a model system with broken inversion symmetry and strong electron-phonon coupling in the quasi-1D regime. 
\end{abstract}

\maketitle

\section{Introduction}

Collective excitations play a crucial role in understanding emergent quantum phenomena and phases in condensed matter physics \cite{Rodin2020,stringari1996collective}. Symmetry or lack of it determines the underlying  dynamics of these collective phenomena and phases\cite{Goldstone1962,Keimer2017,Du2021}. For instance, non-centrosymmetric materials, which lack inversion symmetry, exhibit a range of unique physical phenomena including unconventional superconductivity\cite{Sato2017, Mondal2012,Smidman2017}, Weyl physics\cite{Armitage2018,Morimoto2016}, piezoelectricity\cite{Zhu2015}, and ferroelectricity\cite{Cheong2007,Shi2013}, whereas the presence of symmetry can give rise to  various novel quantum phases with interesting properties\cite{Senthil2015}. For example, topological insulators exhibit bulk band gaps with gapless metallic surface states protected by inversion symmetry \cite{ Qi2008, Fu2007}. Consequently, the discovery of new material with broken inversion symmetry holds great potential for hosting exotic quantum phases. 
%\cite{Xiao2012,Lv2015}. 

The quasi-one-dimensional materials exhibit the fascinating collective phenomena known as charge density waves (CDWs) accompanied by spontaneous rearrangement of the lattice structure \cite{gruner1988dynamics,FavreNicolin2001}. Recently, the family of quasi-1D transition metal tetra chalcogenides, (MX$_4$)$_n$Y with metal (M) = Ta, Nb; Chelcogen (X) = S, Se, or Te, and Halogen (Y) = I, Br) has been the subject of intense research owing to their exotic physical properties arising from the interplay of the single particle excitation and the highly correlated response  of the electronic state \cite{Gressier1984,Gressier1984a,Kwok1989,FavreNicolin2001,TournierColletta2013,Sherwin1984,Gooth2019}. One notable member of this family, (TaSe$_4)_2$I is an enantiomorphic system without any inversion center. It has been reported to be a topological Weyl semimetal, whose Weyl nodes with opposite chirality are coupled by CDW modes resulting in an Axion state \cite{Gooth2019,Li2021,Shi2021,Mu2021}. Concomitantly, in our recent study, (TaSe$_4)_3$I has been found to undergo an inversion symmetry breaking structural phase transition (SPT) at T$_S\sim$ 145~K, driven by the hybridization energy gain due to the off-centric movement of the Ta atoms, which wins over the elastic energy loss \cite{arnab2023nTSI}.   Furthermore, the low-temperature non-centrosymmetric phase exhibits an unusual coexistence of two antagonistic quantum phenomena - superconductivity and weak magnetism below 2.5~K~\cite{arnab2021nTSI}. Considering the significance of the inversion symmetry breaking in low-dimension and unusual phase coexistence, (TaSe$_4)_3$I is a promising platform for studying correlated quantum phenomena and underlying collective lattice dynamics~\cite{SoftPhonon}.

The inversion symmetry-breaking phase transitions are always associated with the emergence of new collective lattice dynamics \cite{Scott1974, Zhang2016, De2021, Liu2018}. Over the past few decades, the phonon modes in various systems are successfully investigated using Raman spectroscopy\cite{Cong2020, Krumhansl1992, SoftPhonon}. The Raman spectroscopy has also been used to investigate phase transition in quasi-1D compounds, (NbSe$_4$)$_3$I \cite{Sekine1988, Dominko2016} and (TaSe$_4)_2$I\cite{Ikari1985}. It has been reported that \cite{Sekine1988} (NbSe$_4)_3$I goes through two structural-phase transitions. The high temperature phase transition is a continuous one with a phonon (soft) branch cross-over. On the other hand, another study \cite{Dominko2016} reported no signature of such crossover of the phonon branches across the transition. They claimed to found soft phonon mode associated with the second order phase transition. However, the phonon modes associated with the structural phases of the quasi-1D compound, (TaSe$_4)_3$I, a isoelectronic counterpart of (NbSe$_4)_3$I, are almost unexplored.

%\textcolor{red}{need critical comment----You could have touched upon symmetry-breaking physics in 1D a bit more from the weak 1D TIs like Bi4I4 with large spin−orbit coupling and the local inversion symmetry breaking}

Here, we present a comprehensive investigation of a unique inversion symmetry-breaking phase transition observed at T$_S\sim$ 141~K \cite{arnab2021nTSI, arnab2023nTSI} of (TaSe$_4)_3$I using in-depth Ramanp scattering experiments and theoretical analysis of phonon mode characteristics. By analyzing temperature-dependent spectra, we have identified eleven Raman active vibration modes for the centrosymmetric (CS) phase, while the non-centrosymmetric (NCS) phase exhibits three additional Raman activ e modes in addition to the previously observed eleven modes. Through a combination of polarized Raman spectroscopy and ${ab}$ ${intio}$ calculations, we have successfully identified all Raman active modes. Among the three newly appeared modes two vibrational modes which were initially Raman inactive at room temperature became Raman active at low temperature due to the absence of inversion symmetry in the crystal structure. And the another one mode at 200.93 cm$^{-1}$ associated with the $B_1$ symmetry which is identified as the soft phonon mode, disappears above the SPT, which suggests the displacive and first-order nature of the phase transition. Additionally, the Fano-shaped asymmetry of the most intense Raman modes confirms the presence of strong electron-phonon coupling. 
Consequently, this study provides a comprehensive understanding of the evolution of phonon modes across the idiosyncratic symmetry inversion transition in low-dimensional systems.
 
\section{Results and Discussions}
\subsection{Characterization - Crystal structure and symmetry}

\begin{figure}[t]
\begin{center}
\includegraphics[width=1\columnwidth]{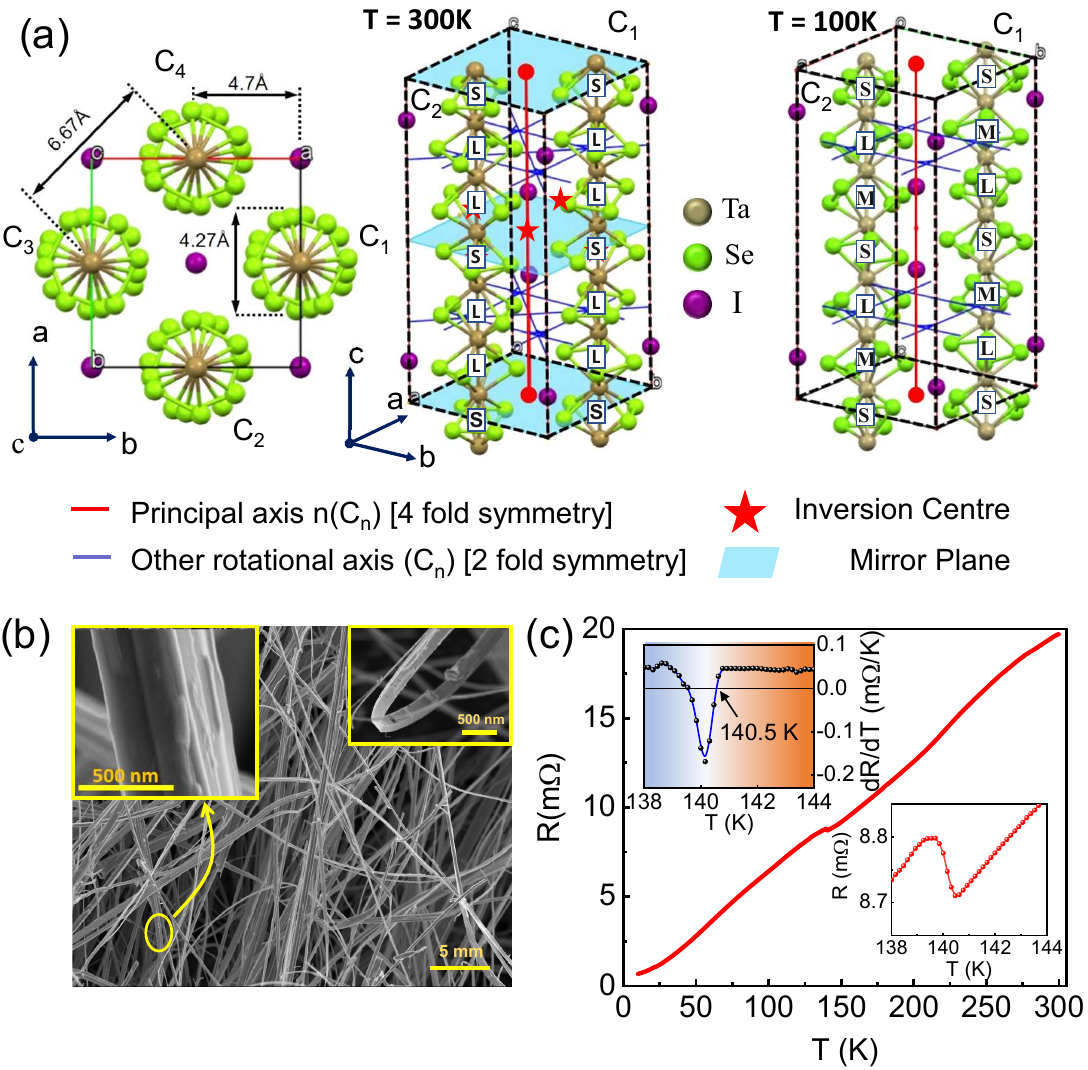}
\caption {\textbf{Temperature induced broken inversion symmetry.} (a) Pictorial representation of crystal structures obtained from SXRD data. The RT structure (at 300~K) shows five centers of inversion, whereas LT  crystal structure (at 100~K) doesn't have any center of inversion. (b)  SEM image of flexible wire-like single crystals of (TaSe$_4)_3$I. (c) The temperature-dependent resistance reveals the inversion symmetry breaking SPT at around T$_S\sim$ 141~K. } 
\label{fig_characterization}
\end{center}
\end{figure}

Ribbon-like single crystals with lengths up to a few millimeters and widths up to a few microns (see Fig.~\ref{fig_characterization} (b)) were grown using the chemical vapor transport method, as described in our earlier works \cite{arnab2021nTSI,arnab2023nTSI} and also briefly in the method section. Single crystal x-ray diffraction (SXRD) analysis revealed a simple tetragonal centrosymmetric crystal structure (space group $P4/mnc$, no. 128) at room temperature. The unit cell has four TaSe$_4$ chains separated by I atoms as shown in Fig.~\ref{fig_characterization} (a). There are 8 Ta atoms of two inequivalent classes Ta(1) and Ta(2) in each chain. The bond length between the Ta-metals shows a sequence along a TaSe$_4$-chain is a repetition of [-Short (3.059\AA)- Long (3.251\AA)-Long (3.251\AA)-] bonds. Whereas the low-temperature crystal structure adopts a non-centrosymmetric configuration, belonging to the same tetragonal crystal family with a space group $P\bar{4}2_1c$ (no.~114). The small lattice distortion changes the bonding sequence of TaSe$_4$ chains and breaks the inversion center \cite{arnab2023nTSI}. The 8 Ta atoms in the chain are split into 3 inequivalent classes Ta(1), Ta(2), and Ta(3). The structural distortion is caused by the shift of the Ta(3) atom towards the Ta(2) atom in chain~1~(C$_1$) and away from Ta(2) atoms in chain~2~(C$_2$). The bonding sequence between two nearest chains are like [ - Long(L) (3.306\,\AA) - Medium(M) (3.116\,\AA) - Short(S) (3.051\,\AA) - ] and [ - Long(L) (3.306\,\AA) - Short(S) (3.051\,\AA) - Medium(M) (3.116\,\AA) - ]. The shift of the Ta(3) atom is in the reverse direction for the two nearest neighbor chains making it non-polar. 

The CS structure has five inversion centers (IC), with four located at the center of each TaSe$_4$ chain and the remaining one at the center of the ${Tetragonal}$ unit cell. However, there is no inversion center in the low-temperature crystal structure as shown in Fig.~\ref{fig_characterization}(a). The inversion symmetry breaking SPT is also reflected as an anomaly at T$_S\sim$ 141~K in the temperature-dependent resistance as shown in Fig.~\ref{fig_characterization}(c). Notably, the lower inset along with d$R$/d$T$ vs. T plot in the upper inset clearly reveals the above structural transition~\cite{arnab2023nTSI}.

%The inversion symmetry breaking structural phase transition is also reflected as an anomaly at T$_s$=141~K in the temperature-dependent resistivity (see Fig.~\ref{fig_characterization}(c)) \textcolor{red}{add few lines related to the observation in resistivity}. \textcolor{blue}{In the d$R$/d$T$ vs T plot (see upper inset) we have observed negative value of d$R$/d$T$ near the structural transition region. We define the structural transition temperature at which the resistance first starts to increase and it is corresponds to the point at which the dR/dt initailly crosses 0.}

\subsection{Raman signatures of structural phase transition.}
\begin{figure*}[t]
\begin{center}
\includegraphics[width=2\columnwidth]{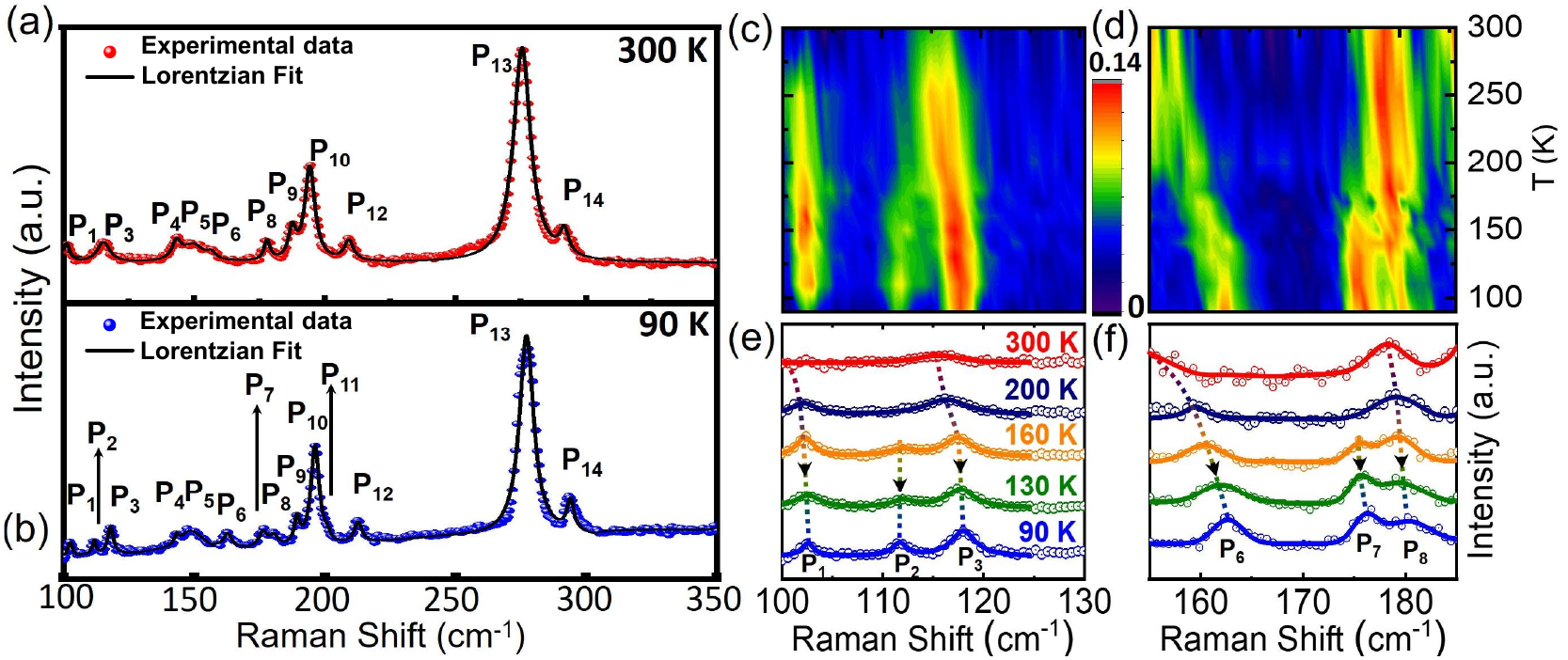}
\caption {\textbf{Raman spectra of (TaSe$_4$)$_3$I.} (a-b) Representative Raman spectra at 300 K and 90 K. Black solid lines represent the total fit with a sum of Lorentzian functions to the experimental data, presented in colored scattered points. (c-d) Temperature evolution of Raman spectra in the specific spectral region of interest in 2D contour plots. (e-f) Temperature evolution of specifically selected Raman active phonon modes, P$_1$, P$_2$, P$_3$, P$_6$, P$_7$, and P$_8$ across the phase transition at around T$_S\sim$ 141~K. The dashed line is a guide to the eye for better visualization of the shift of the phonon modes.} 
\label{fig_ramancontour}
\end{center}
\end{figure*}

\begin{table*}
\caption{\label{raman_theory} \textbf{Details of the Raman active modes}. The experimental (Exp.) and calculated (Theo.) Raman active vibration modes of (TaSe$_4$)$_3$I at 90~K and 300~K. Origin describes the vibration among elements responsible for the corresponding Raman active modes.}
%\centering
\begin{tabular}{ |p{1.1cm}|p{2.1cm}|p{2.1cm}|p{1.1cm}|p{1.3cm}|p{2.1cm}|p{2.1cm}|p{1.1cm}|p{1.3cm}| }
%\hspace{-1cm} 
 \hline
  \multirow{2}{0em}{Raman modes}&\multicolumn{4}{|c|}{CS phase}&\multicolumn{4}{|c|}{NCS phase} \\
    \cline{2-9}
     & Exp. & Theo. & Symm. & Origin & Exp. & Theo. & Symm. & Origin\\

\hline
P$_1$ &101.58 $cm^{-1}$ & 101.97 $cm^{-1}$ & $A_{1g}$ & Se-Se &102.44 $cm^{-1}$ & 101.40 $cm^{-1}$ & $A_{1}$& Se-Se \\
\hline
P$_2$ & ${Absent}$ & ${Absent}$ & $E_u$ & Ta-Se-I& 111.53 $cm^{-1}$ & 110.31 $cm^{-1}$ & $E$& Ta-Se-I\\
\hline
P$_3$ & 115.39 $cm^{-1}$ & 116.21 $cm^{-1}$ & $B_{1g}$ & Ta-Se& 117.90 $cm^{-1}$ & 116.88 $cm^{-1}$ & $B_1$& Ta-Se\\
\hline
P$_4$ & 143.19 $cm^{-1}$ & 141.36 $cm^{-1}$ & $A_{1g}$ & Se-Se & 143.31 $cm^{-1}$ & 144.16 $cm^{-1}$ & $A_1$& Se-Se\\
\hline
P$_5$ & 148.42 $cm^{-1}$ & 148.33 $cm^{-1}$ & $B_{1g}$ & Se-Se & 149.73 $cm^{-1}$ & 150.60 $cm^{-1}$ & $B_{1}$ & Se-Se\\
\hline
P$_6$ & 156.50 $cm^{-1}$ & 158.14 $cm^{-1}$ & $E_{g}$ & Ta-Se & 162.82 $cm^{-1}$ & 161.81 $cm^{-1}$ & $E$ & Ta-Se \\
\hline
P$_7$ & ${Absent}$ & ${Absent}$ & $E_u$ & Ta-Se & 176.40 $cm^{-1}$ &175.62 $cm^{-1}$ & $E$ & Ta-Se\\
\hline
P$_8$ & 178.42 $cm^{-1}$ &178.35 $cm^{-1}$ & $B_{2g}$ & Se-Se & 180.56 $cm^{-1}$ & 180.96 $cm^{-1}$ & $B_2$ & Se-Se\\
\hline
P$_9$ & 187.47 $cm^{-1}$ & 189.09 $cm^{-1}$ & $B_{1g}$ & Ta-Se & 189.57 $cm^{-1}$ &190.86 $cm^{-1}$ &$B_1$ & Ta-Se\\
\hline
P$_{10}$ & 193.80 $cm^{-1}$&194.10 $cm^{-1}$ & $A_{1g}$ & Se-Se & 196.18 $cm^{-1}$ &194.47 $cm^{-1}$ & $A_1$ & Se-Se\\
\hline
P$_{11}$ & ${Absent}$ & ${Absent}$ & ${Absent}$ & ${Absent}$ & 200.93 $cm^{-1}$ & 197.40 $cm^{-1}$ & $B_1$ & Ta-Se\\
\hline
P$_{12}$ & 209.14 $cm^{-1}$&208.88 $cm^{-1}$& $E_g$ & Ta-Se & 212.67 $cm^{-1}$ &209.88 $cm^{-1}$ & $E$ & Ta-Se\\
\hline
P$_{13}$ & 275.45 $cm^{-1}$&285.13$cm^{-1}$ & $A_{1g}$ & Ta-Se & 277.14 $cm^{-1}$ & 284.59 $cm^{-1}$ & $A_1$ & Ta-Se\\
\hline
P$_{14}$ & 291.72 $cm^{-1}$& 301.14 $cm^{-1}$ & $B_{1g}$ & Ta-Se & 293.86 $cm^{-1}$ &302.94 $cm^{-1}$&$B_1$ & Ta-Se\\
\hline
\end{tabular}
\end{table*}

To gain deeper insights into the structural phase transition, temperature-dependent Raman scattering experiments have been carried out on a bulk Crystal of (TaSe$_4)_3$I in the temperature range 90~K to 300~K (see raw data in the supplementary section, Fig.~S2). Fig.~\ref{fig_ramancontour}(a,b) depicts the representative Raman spectra in the spectral range of 100 cm$^{-1}$ $<$ $\omega$ $<$ 350 cm$^{-1}$ at 90~K (LT) and 300~K (RT), respectively. The Raman spectra are fitted with a sum of Lorentzian functions to determine the peak frequency ($\omega$) and line width (FWHM).

The Raman spectrum at 300~K (see Fig.~\ref{fig_ramancontour}(a)) reveal eleven Raman active vibration modes at approximately, 101.58(P$_1$), 115.39(P$_3$), 143.19(P$_4$), 148.42(P$_5$), 156.50(P$_6$), 178.42(P$_8$), 187.47(P$_9$), 193.80(P$_{10}$), 209.14(P$_{12}$), 275.45(P$_{13}$), and 291.72(P$_{14}$) cm$^{-1}$.  Whereas, Raman spectrum at 90~K shows three additional peaks at 111.53 (P$_2$), 176.40 (P$_7$), and 200.93 (P$_{11}$) cm$^{-1}$ (see Fig.~\ref{fig_ramancontour}(b-f)) alongside the modes observed at room temperature. The details of these identified vibration modes, including their mode of symmetry, and origin, are summarized in Table~\ref{raman_theory}. The emergence of these additional Raman-active modes below T$_S$ can be attributed to the SPT accompanied by a symmetry-breaking effect \cite{Zhang2016,Gao2018}, as discussed later.

\begin{figure*}
\begin{center}
\includegraphics[width=2\columnwidth]{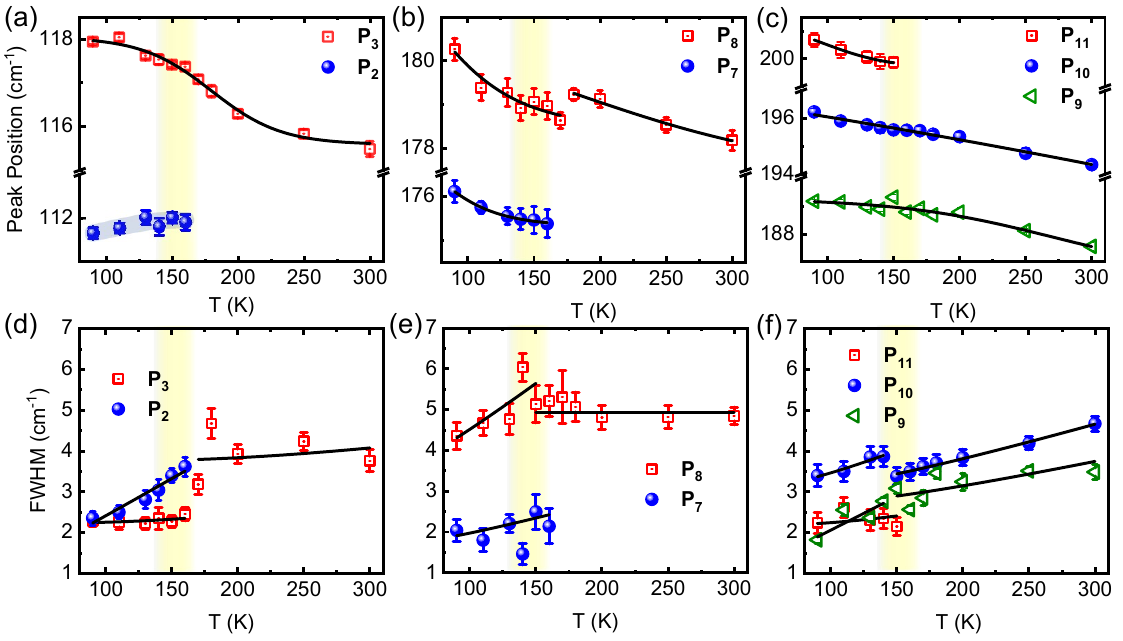}
\caption { \textbf{Thermal evolution of characteristics parameters of Raman active modes.} Temperature evolution of the self-energy parameter and FWHM for a few selected Raman modes of special interest. Scattered red, blue, and green symbols are experimental data, whereas the continuous black lines are fitted using the symmetric three-phonon coupling model. The yellow-shaded regions represent the temperature regime of phase transition.}
\label{Thermal_evolution}
\end{center}
\end{figure*}

%%%%% %%%%%%%%%%%%%%%%%%%%%%%%

%\sout{For better clarity of the intriguing signature of the phase transition in Raman spectra, we have estimated the mode frequencies,line-width or full widths at half maximum (FWHM), and peak intensity as a function of temperature from the temperature-dependent Raman spectra (see Fig.~Sxx in Supplementary). The frequencies of specifically selected  vibration modes (P$_2$, P$_3$, P$_6$, P$_7$, P$_8$, P$_{9}$, and P$_{10}$) are presented in Fig.~\ref{Thermal_evolution}. The remaining characteristic parameters of Raman modes are included in the supplementary section (see Fig.~SX).}

%Except, P$_2$ all other Raman active modes shift towards higher frequencies (${\omega}$) with decreasing temperature.

%due to the suppression of anharmonic effects at lower temperatures (see P$_3$,  P$_5$, and P$_7$ in Fig.~\ref{fig_ramancontour} and also Fig.~SXX in supplementary).

\subsection{Temperature evolution of Raman active modes.} Signatures of structural transition are clearly visible in the temperature dependence of Raman spectra. With decreasing temperature, three additional Raman modes (P$_2$, P$_7$, and P$_{11}$) appear near the transition suggesting their association with the symmetry of crystal structure as shown in Fig.~\ref{Thermal_evolution}. At around T$_S$, one Raman active mode (labeled as P$_7$) show discrete jumps in the mode frequencies. 
Except for P$_2$ (discussed later), all the Raman modes show blue shifts, i.e., mode frequency ($\omega$) increases with decreasing temperature (\textcolor{blue}{see Fig.~\ref{Thermal_evolution}}). Since the phonon modes involve the vibrations of different atoms with varying degrees as well as directions, the amount of the blue shift in each individual mode occurs at different rates and slightly different temperatures. The temperature evolution of Raman active vibration modes is analyzed using the symmetric three-phonon coupling phenomenological model \cite{Fitting_Raman},

\begin{equation}
    \omega(T) = \omega_1 + \frac{\omega^\prime - \omega_1}{1+exp\frac{T-T_0}{dT}}
\end{equation}

Here, $\omega^\prime$ and $\omega_1$ represent the top and bottom of the sigmoidal curve, respectively. T$_0$ is the center point and d$T$ controls the width of the curve. The results are well described in Fig.~\ref{Thermal_evolution}. 

In contrast, the  P$_2$ mode shows anomalous redshift, i.e., ${\omega}$ decreases with decreasing temperature. This anomalous behavior of P$_2$ has been considered as the possible reflection of strong anharmonic phonon-phonon interactions mediated through structural distortion in Ta-chain \cite{AnharmonicPhonon}. 

The modes become sharper  with decreasing temperature,
as shown in Fig.~\ref{Thermal_evolution}(d-f) and Fig.~S5. The line widths, i.e., The full width at half maximum (FWHMs), monotonically decrease with decreasing temperature because of increasing phonon lifetime \cite{PhononLifetime}. Interestingly, the FWHMs demonstrate abrupt step-like changes across the transition temperature at T$_S$, indicating a distinct change in the coupling between the lattice and phonon degrees of freedom \cite{Gao2018}. The temperature-dependent line width ($\Gamma$) is described by symmetric three-phonon coupling models \cite{Fitting_Raman}, 
\begin{equation}
    \Gamma = \Gamma_0\left(1 + \frac{1}{exp\frac{hc\bar{\nu_1}}{k_BT}-1} + \frac{1}{exp\frac{hc\bar{\nu_2}}{k_BT}-1}\right)
\end{equation}
Here, $h$ and $k_B$ are the Planck and Boltzmann constants, respectively. The $c$ is the speed of light, and T is the temperature. The $\Gamma_0$ is the asymptotic value of the linewidth at zero temperature. $\bar{\nu_1}$ and $\bar{\nu_2}$ are two acoustic phonon modes with different wavenumbers and opposite wavevectors. The observed changes in the characteristics of Raman modes across the phase transition, coupled with the emergence of three new active Raman vibration modes in the low-temperature spectra, provide direct evidence of a SPT occurring at T$_S$.

\begin{figure}
\begin{center}
\includegraphics[width=1\columnwidth]{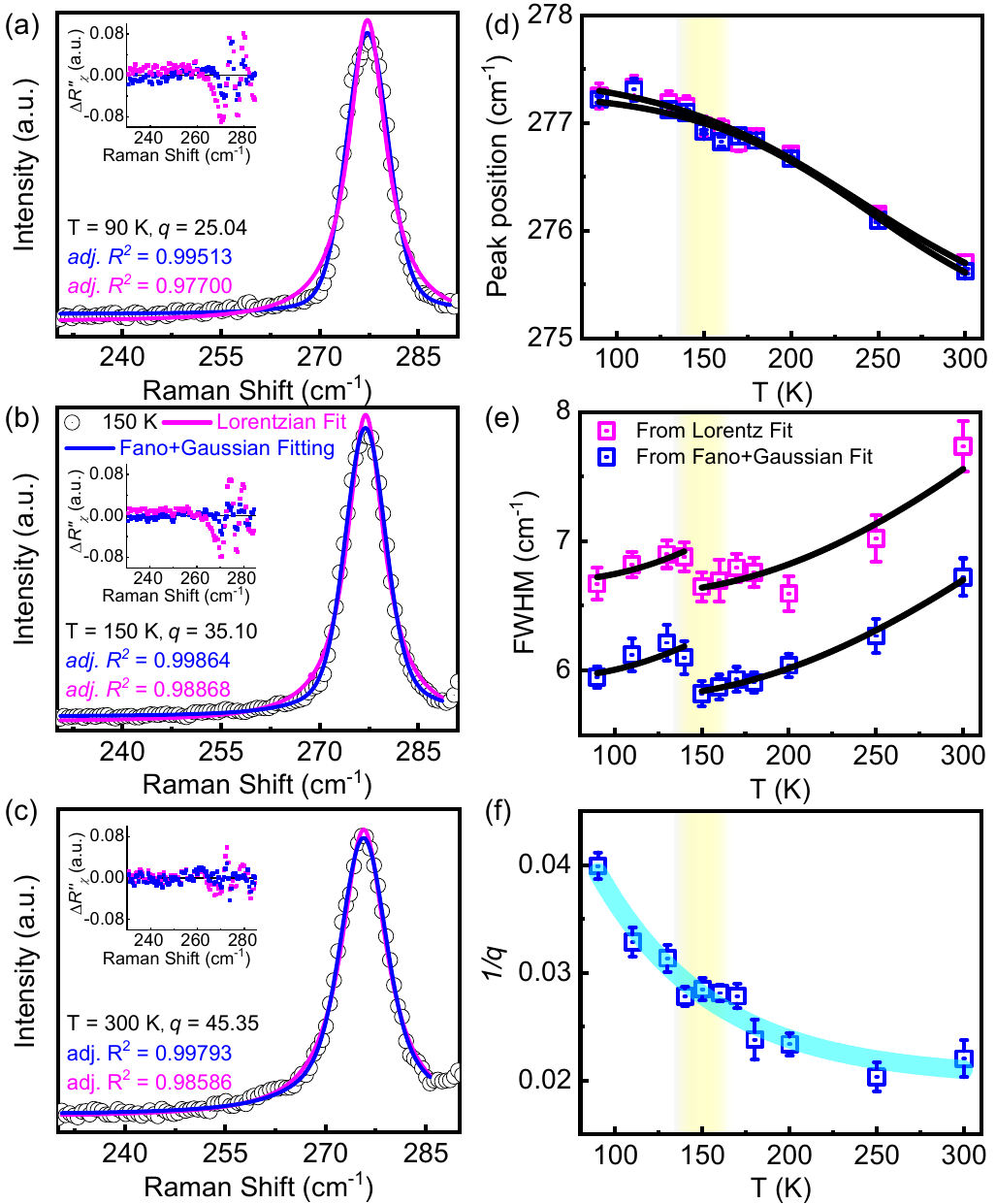}
\caption {\textbf{Fano asymmetry: Electron-phonon coupling.} (a-c) Fano line shape analysis of the P$_{13}$ Raman mode. The blue solid lines depict the fit obtained by convolving the Fano profile with a Gaussian, while the magenta solid lines represent Lorentzian profiles. The insets exhibit residual plots, highlighting the deviations between the fitted profiles and the experimental data. Temperature dependence of the extracted parameters (d) peak position and (e) line width (FWHM) from Fano-Gaussian fit and Lorentzian fit is presented. The solid black line corresponds to the theoretical fit. (f) presents the temperature-dependent evolution of the inverse of the Fano asymmetry parameter i.e. $1/q$ vs T.}
\label{Fano}
\end{center}
\end{figure}

 %%%%% %%%%%%%%%%%%%%%%%%%%%%%%

\subsection{Fano asymmetry of Raman spectra: Electron-phonon coupling}

Closer examination of the Raman spectra reveals finer intriguing characteristics of Raman spectra. Notably, the most intense peak, P$_{13}$ clearly exhibits Fano-shaped asymmetry, which can be attributed to the coupling between the discrete phonon spectra and  electronic scattering continuum \cite{Hasdeo2014, DjurdjifmmodecuteclseciMijin2023}. To gain a deeper understanding, P$_{13}$ mode is analyzed using the Fano function, $  I(\omega)=I_0\frac{(q+\epsilon)^2}{1+\epsilon^2}$ \cite{Fano1961}. Here,  ${\epsilon}$(${\omega}$) = 2(${\omega}$-{${\omega}_0$})/${\Gamma}$ where ${\omega}_0$ is the phonon frequency in the absence of interaction, $\Gamma$ is the FWHM, and I$_0$ is the intensity. 
The parameter $q$ represents the "Fano" parameter that describes the asymmetry. Additionally, the electron-phonon coupling strength can be inferred from the inverse of the Fano parameter i.e. $1/|q|$ \cite{FanoZhang2015,DjurdjifmmodecuteclseciMijin2023}. To account for the finite spectral resolution, the P$_{13}$ peak is fitted using the aforementioned Fano profile convolved with a Gaussian function~\cite{FanoZhang2015}. The comparison between the model fits of the experimental data is presented in Fig.~\ref{Fano}(d,e) displays the temperature variation of the phonon frequency ${\omega}_0$, the FWHM and the electron-phonon coupling constant, $1/|q|$ for the P$_{13}$ mode. The exponential decay of $1/|q|$ with temperature signifies that the electron-phonon coupling strength weakens at higher temperatures.

%\textcolor{red}{need little more explanation!}

\begin{figure*}[t]
\begin{center}
\includegraphics[width=2\columnwidth]{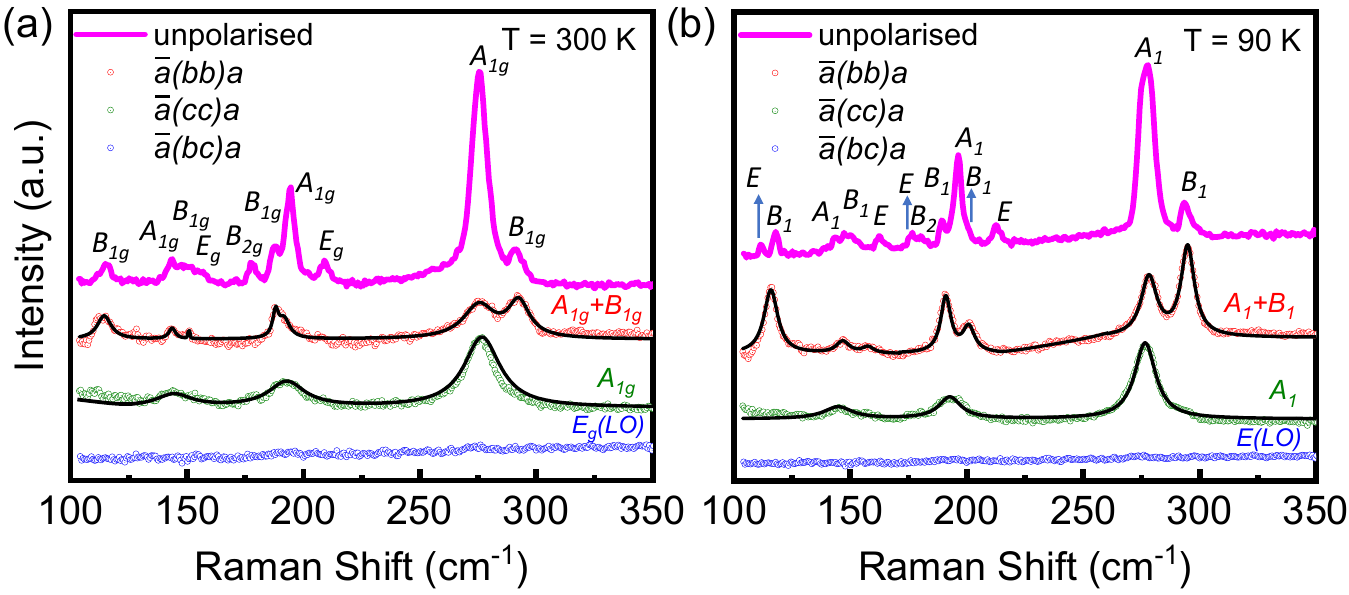}
 \caption {\textbf{Polarised Raman spectra.} Polarized Raman spectra were measured at (a) 300~K and (b) 90~K. The identified Raman peaks are labeled with reference to their associated symmetry and selection rule.}
\label{polariastion_resolved}\end{center}
\end{figure*}

\subsection{Identification of Raman active modes}

To gain further insights into these Raman modes, such as their symmetry identification and origin, we conducted a group theory analysis of the vibration modes summarized in table~\ref{raman_theory}. 

\textbf{Raman active modes  for centrosymmetric structure.} The theoretical analysis of the vibration modes for a tetragonal crystal structure with space group: $P4/mnc$ is carried out  using lattice dynamics calculations. Since there are 64 ($n$) atoms in the unit cell for both structures, the vibrational modes in the centrosymmetric phase differentiate into 192 (3$n$) mechanical representations,$\Gamma_{D4h} = \Gamma_{D4h}^{optic} + \Gamma_{D4h}^{acoustic}$. Here, three modes are acoustic modes given by, ($\Gamma_{D4h}^{acoustic}$ = $A_{2u}$ + $E_u$). The remaining 189 modes are optical as given by, ($\Gamma_{D4h}^{optic}$ = 12$A_{1g}$ + 10$A_{1u}$ + 13$A_{2g}$ + 10$A_{2u}$ + 12$B_{1g}$ + 10$B_{1u}$ + 11$B_{2g}$ + 9$B_{2u}$ + 27$E_u$ + 24$E_g$). Among these optical modes, (10$A_{2u}$ + 27$E_u$) are infrared(IR) active and Raman -inactive modes, while (12$A_{1g}$ + 12$B_{1g}$ + 11$B_{2g}$ + 24$E_g$) shows the reverse characteristics i.e. Raman active, but IR inactive. 

\textbf{Raman active modes for non-centrosymmetric structure.} Similarly, the vibrational modes in the non-centrosymmetric phase can be decomposed into 192 mechanical representations, denoted as $\Gamma_{D2d} = (B_2 + E) + (22A_1 + 22A_2 + 22B_1 + 21B_2 + 51E)$. Here, the first and second groups of the representation correspond to the acoustic and optical modes. Among optical modes, the IR-active and Raman active modes are  21$B_2$ + 51$E$, and  22$A_1$ + 22$B_1$ + 21$B_2$ + 51$E$, respectively. Although the total number of vibration modes remains the same, the number of Raman-active modes in the non-centrosymmetric phase (total 167 modes) is found to be much greater than the centrosymmetric phase (total 83 modes). This reflects in the appearance of additional Raman modes in low-temperature Raman spectra as shown in Fig.~\ref{fig_ramancontour}(a,b), which is in very good agreement with the simulated modes presented in Table~\ref{raman_theory} and described in details in the supplementary section.

\subsection{Polarised Raman spectra and Porto's notation} 
The symmetries of the Raman active modes are identified by polarization-dependent Raman scattering experiments \cite{Porto}. Based on the theoretical analysis, the selection rules are shown in Table \ref{tab:widgets} with the help of Porto's notation (P.N.) backscattering geometry (details can be found in the supplementary section S4). Corresponding to $P4/mnc$ space group, the $A_{1g}$ mode is allowed in the $\overline{a}(bb)a$ and $\overline{a}(cc)a$ configurations, while the $B_{1g}$ modes along $\overline{a}(bb)a$ configuration. The $E_g$(${LO}$) mode is observable in the $\overline{a}(bc)a$ configuration. The experimental setup was designed to access the vibrational modes listed in Porto's notation.  To ensure accurate mode assignments based on the selection rules, the polarization of the incident and scattered photons was carefully controlled using a polarizer and analyzer, respectively. 
\begin{table*}
\caption{\label{tab:widgets} \textbf{Selection rule for polarization dependent Raman scattering.} Back scattering geometry (Y represents the modes that can be observed in each one of the directions and N represents the mode that can not be observed.)}
\begin{center}
\begin{tabular}{ |p{2cm}|p{1.5cm}|p{1.5cm}|p{1.5cm}|p{1.5cm}|p{1.5cm}|p{1.5cm}|p{1.5cm}|p{1.5cm}| }
\hline
\multicolumn{1}{|c|}{Porto's Notiation}&\multicolumn{4}{|c|}{Space Group $P4/mnc$ no 117 (RT)}& \multicolumn{4}{|c|}{Space Group $P\overline{4}2_1c$ no 114 (LT)} \\
\hline
& $A_{1g}$ & $B_{1g}$ & $B_{2g}$(${LO}$) & $E_g$(${LO}$) & $A_1$ & $B_1$ & $B_2$(${LO}$) & $E$(${LO}$)\\
\hline
$\overline{a}(bb)a$ & Y & Y & N & N & Y & Y & N & N\\
\hline
$\overline{a}(bc)a$ & N & N & N & Y & N & N & N & Y \\
\hline
$\overline{a}(cc)a$ & Y & N & N & N & Y & N & N & N \\
\hline
\end{tabular}
\end{center}
\end{table*}

%%%%% %%%%%%%%%%%%%%%%%%%%%%%%

\begin{figure*}[t]
\begin{center}
\includegraphics[width=2\columnwidth]{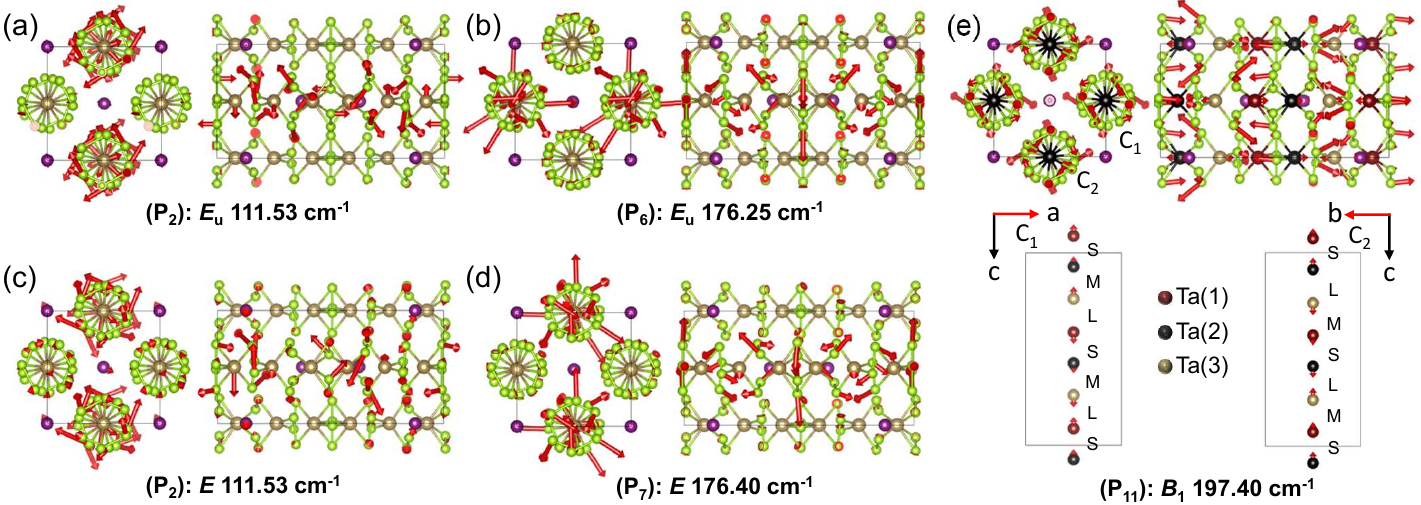}
        \caption {\textbf{Simulated patterns of specifically selected vibrational modes.} Calculated (a,b) vibrational patterns (IR active only) in the centrosymmetric structure, and (c,d) Raman active modes in the non-centrosymmetric phase corresponding to P$_2$, and P$_7$ modes, which are directly sensitive to inversion symmetry breaking structural transition. The arrows scale the atomic vibration direction. (e) Vibration pattern corresponding to the soft Raman mode, P$_{11}$, which disappears for the centrosymmetric structure above T$_{S}$.}
	\label{fig_VibrationalMode_1}
	\end{center}
\end{figure*}

The sample was aligned in such a way that its crystal axis direction was along the propagation of the incident laser. Consequently, the incident laser direction ($k_i$) corresponds to $\overline{a}$, while the scattered laser direction ($k_s$) corresponds to $a$.To control the polarization direction of the incident light ($E_i$), we utilized a $\lambda$/2 plate and selected the appropriate configuration. Fig.~\ref{polariastion_resolved}(a,b) summarizes the polarized Raman spectra measured at both 100~K and 300~K. The Raman spectra exhibit seven peaks in the $\overline{a}(bb)a$ combination originated from the $A_{1g}$ and $B_{1g}$ vibrational modes. By contrast, the $\overline{a}(bc)a$ configuration exhibited three vibrational modes of $A_{1g}$, allowing us to distinguish between the two modes. No modes were observed in the $\overline{a}(bc)a$ configuration, preventing us from identifying any $E_g$(LO) modes within our experimental range. However, $E_g$(TO) mode can be present. In that case, a right-angle Raman scattering geometry is needed which is beyond our experimental limit. 

Therefore, our polarization-dependent study reveals three $A_{1g}$ and four $B_{1g}$ modes in HT spectra, whereas, in LT (90 K) spectra, we can identify three $A_1$ and five $B_1$ using Porto's configuration.  The newly appeared $B_1$ mode in LT Raman spectra corresponds to the P$_{11}$ mode. Further details regarding the nature of this P$_{11}$ mode are discussed later.

\subsection{Vibration patterns of Raman modes}

The vibration patterns of Raman modes are simulated for two different crystal symmetries.  Fig.~\ref{fig_VibrationalMode_1}(a-d) depicts the simulated vibrational patterns that are sensitive to breaking the crystal structure's inversion symmetry. The simulated vibration patterns of the remaining modes are presented in Fig.~S10 and S11 in the supplementary section. 

%The spectroscopic selection rule \cite{D.A.Long_book} based on lattice symmetry tells us, the Raman and IR-active modes are found to be mutually exclusive in the centrosymmetric atmosphere, while they may appear simultaneously in a situation when there is no inversion symmetry point. 

%Our as a studied compound, (TaSe$_4$)$_3$I undergoes a unique SPT at around 150 K; RT centrosymmetric to LT non-centrosymmetric phase \textcolor{red}{[R]}. Now, in accordance with \cite{D.A.Long_book}, here Raman in-active A$_2u$ and $E_u$ modes  evolve to A$_2$ and $E$ (which are both Raman and infrared active) with reducing T from RT, below SPT, providing direct affirmation on symmetry breaking.

\subsubsection{Signature of broken inversion symmetry.}
Three additional Raman modes P$_2$, P$_7$, and P$_{11}$ appear below T$_S$ as shown in Fig.~\ref{Thermal_evolution}. Symmetry analysis reveals that P$_2$ and P$_7$ belong to the identical irreducible representation of $E$ for NC structure, and their corresponding vibration patterns are depicted in Fig.~\ref{fig_VibrationalMode_1}(c,d). Though the vibrational patterns of these modes i.e. P$_2$ and P$_7$ above T$_S$ remain the same as shown in Fig.~\ref{fig_VibrationalMode_1} (a-d), the corresponding representation of $E$ becomes $E_u$ for the crystal structure with an inversion center. As there is inversion symmetry in the RT phase then all the lattice vibrations at the ${\Gamma}$ point have either even or odd parity with respect to the inversion center, and the odd ones are Raman-inactive due to selection rules \cite{D.A.Long_book}. Hence, this vibrational mode $E_u$ is antisymmetric (i.e. odd parity) with respect to the inversion center and these modes P$_2$ and P$_7$ are now Raman inactive. 
Therefore, the modes are absent in Raman spectra above T$_S$. Hence, the presence of these modes (P$_2$ and P$_7$) below T$_S$ provides direct affirmation of the inversion symmetry-breaking transition in (TaSe$_4)_3$I.
\subsubsection{Softening of phonon mode,  P$_{11}$}

The P$_{11}$ Raman peak belonging to $B_1$ symmetry, coexists with nearest peaks P$_{10}$ and P$_9$ at low temperatures as shown in Fig.~\ref{fig_ramancontour}(a) and Fig.~S3.  The temperature evolution of the  P$_9$, P$_{10}$, and P$_{11}$ Raman peaks shows that these modes shift to lower frequencies with increasing temperature and the line widths become broader as shown in Fig.~\ref{Thermal_evolution}. As shown in Fig.~S3 the intensity of P$_{11}$ gradually decreases, whereas the intensity of the other two peaks increases with increasing temperature. The behavior of these phonon modes can be explained by the following model. At the ${\Gamma}$ point, the  P$_{11}$ is a soft phonon mode associated with the structural transition, whereas P$_9$ and P$_{10}$  are not sensitive to the structural transition. With increasing temperature, the frequency of the soft mode approaches that of P$_9$ and P$_{10}$ modes and vanishes above the phase transition. \\
%The  P$_{10}$ is a soft phonon at the ${\Gamma}$ point associated with the structural transition. Moreover, there are other phonons P$_8$ and P$_9$  at the ${\Gamma}$ point with a fixed frequency, which is not associated with the phase transition. In the low-temperature phase, the frequency of the soft mode approaches that of the P$_8$ and P$_9$ modes and gradually vanishes above the transition. 
Fig.~\ref{fig_VibrationalMode_1}(e) shows the simulated vibrational pattern of the P$_{11}$ mode with $B_1$ symmetry, that is antisymmetry with respect to the principle rotational axis and symmetry with respect to other rotational axes. To illustrate this, we have presented the vibrational pattern of Ta atoms of the two nearest chains in Fig.~\ref{fig_VibrationalMode_1}(e). The Ta atoms vibrate along the chains. 
%If we closely look at the chain C$_1$, the vibrational direction of Ta(3) atom is towards the Ta(2) atom and opposite to Ta(1) atom. Consequently, Ta(3) moves closer to Ta(2), resulting in a shorter bond length (Medium), whereas Ta(3) moves away from Ta(2), leading to a longer bond length (Long). This explains the bonding sequence of C$_1$ which is -[S-M-L]-. On the other hand, for chain C$2$, the vibrational direction of Ta(3) is toward the Ta(1) atom and opposite to the Ta(2) atom, resulting in a bonding sequence of -[S-L-M]-.
The movement of Ta(3) atom at low temperature phase change the bond length sequence from -[S-L-L]- to -[S-M-L]-. 
Thus, the displacement of Ta atoms is the primary cause of the structural transition in (TaSe$_4)_3$I. The disappearance of the P$_{11}$ phonon mode can be attributed to the displacive nature of this structural transition. 

\subsection{Theoretical calculations: Nature of phase transition}

 The symmetry of the phonon-modes is used to trace the Raman active phonon in the LT and RT phases of (TaSe$_4)_3$I. RT phase of (TaSe$_4)_3$I, being centrosymmetric, have phonon modes that are either of even parity or odd parity. It is found that only odd parity phonon modes are Raman-inactive by applying selection rules. Experimental results show that new phonon modes with the vibrational configuration are present as Raman-inactive modes in the RT phase but become Raman-active in the LT phase. By inspection of the modes, we found that newly arose modes are of odd parity and therefore were not present in RT but present in the LT phase, as it lacks in having inversion point, and the selection rule does allow it to be Raman-active. The selection rule in both phases can be understood by analyzing the character table of the symmetries of the phases (see supplementary section S4).

We can find the nature of the phase transition by tracing the phonon modes with changing temperatures. We have traced each intensity peak in Fig.~S2 to guide us in finding the phonon modes that are decreasing in amplitude gradually with the increasing temperature and disappearing near transition temperature (T$_S$). These modes are called soft-phonons \cite{SoftPhonon}, and the process mentioned is called softening of the phonon modes. The calculated frequencies of the soft phonons above the transition temperature are imaginary. It signifies that the oscillation of atoms relating soft modes seized to oscillate as there is no restoring force regarding the old equilibrium position, and the force generated from small oscillation with respect to an old equilibrium point shifts the atom further till the new equilibrium position is reached. The starting and final structures in this process differ in structure and symmetry.
The above process can solely be driven by temperature alone and described by phenomenological modeling using Landau's theory of phase-transition of the second kind. Apart from dynamics, SPT can be of first order with the sudden disappearance of phonon mode \cite{Waghmare1997, Krumhansl1992, Krumhansl1990Aug, Krumhansl1989Feb}.

various complex processes can drive the softening of the phonon. One such scenario is when the electron-phonon coupling is of significant strength to renormalize the bare phonon frequencies and thus soften them. The resultant dip in the phonon branch is called the Kohn anomaly. If the electron-phonon coupling strength, along with electron-electron correlation, becomes significant enough to reduce the value of the phonon frequency to zero, it can induce a SPT by the process mentioned earlier. Peierls transition in 1D metal \cite{js2,cdw_mis,nesting} goes through each step to form a charge density wave(CDW)insulator\cite{ph_soft_Cr,optical_soft,Hoesch2009}. Theoretical calculation of the temperature dependency of the Lindhard function can shed light on this transition. In quasi 1D-metals, such transitions are well reported and assigned to sharp peaks in charge-charge correlation or Lindhard function\cite{nesting} as an indicator of charge correlation. It implies that charges can rearrange themselves with little or no outside potential as impetus. Though a sharp peak in charge-charge correlation alone can induce a commensurate or in-commensurate charge distribution by rearranging the electronic charge, but may not result in affecting the crystal structure. Lacking electron-phonon coupling can be the reason for such transitions. The transitions may not follow every cause-effect chain in real metals as in the 1D Peierls model. 

Lack of gradually vanishing modes in Raman spectra along with the experimentally determined arbitrary displacement direction (specifically, along the c-axis), as calculated in the ISOTROPY suite\cite{isotropy_suite, group-subg_phase} for possible second-order transition, help to assign the transition to be first order. Indeed the disappearance of P$_{11}$ mode indicates first-order phase transition. To model the free energy functional for first-order phase transition, we take displacement($d$) of the Ta$^{+5}$ ion (Ta(3)in Fig-7) along the Ta-Ta chain as our order parameter. The second-order phase transition is clearly an event of dynamical instability originating from the vanishing soft phonon and therefore vanishing restoring force. Although it is not the only way to have a displacive phase transition. A first-order phase transition may occur because of thermodynamic instability\cite{Waghmare1997, Krumhansl1992}. Free energy expression, in our case,
\begin{equation} 
F=\frac{1}{2}\lambda d^2+\frac{1}{4} \left(\delta -\frac{\alpha  \beta ^2}{2}\right)d^4+\frac{1}{6}\gamma d^6
\end{equation} 
$d$: displacement of Ta(3) atom, $\beta$: stain coupling (thorough discussion on the choice of functional in supplementary). The sign of the quartic term in the above plays a crucial role in determining the nature of phase transition. The above expression is constructed from invariant monomials~\cite{gtpack1,gtpack2}(listed in supplement) of point group(D$_{4h}$) of the RT phase crystal, therefore invariant under the action of every element of D$_{4h}$. As we can see, the coefficient of the anharmonic  quartic term plays a central role. For the stability argument, we have added $\text{d}^6$ term, the physics played by anharmonicity of a phonon(phonon-phonon interaction) is of a substantial amount. Blue shifting of peak P$_2$ mode confirms it as shown Fig.~\ref{Thermal_evolution}.

The coefficient of the quartic term consists of strain coupling parameters. Without strain coupled with the displacive transition, the nature of it will be of second order\cite{Krumhansl1992}. The Fermi surface of the LT structure is calculated (without relaxation) and shows the presence of nesting. The competitive interplay of charge ordering with other events occurring during displacement can cause the strain coupling with the transition by the reason mentioned above. By analyzing the experimental data and comparing it with first-principles calculations, it is found highly intriguing events such as strain, forces acting against possible charge order, strong electron-phonon coupling, and anharmonicity  playing in a single arena induce such phase transition. 

%It needs a more complete model that encompasses all the agents acting during the transition process.

\section{Summary and Conclusion}

Our study provides compelling evidence of inversion symmetry breaking in the quasi-1D system (TaSe$_4$)$_3$I based on the temperature-dependent evolution of phonon mode characteristics. The unique structural distortion of the Ta-chains leads to the breaking of inversion symmetry at T$_S \sim$ 141~K, as evidenced by the thermal variation of self-energy parameters of phonon modes. Through polarization-dependent Raman scattering combined with theoretical analysis, we have successfully identified two distinct Raman modes (P$_2$ and P$_7$) that unambiguously confirm the absence of inversion symmetry in the low-temperature crystal structure. Additionally, the identification of the P$_{11}$ mode as a soft phonon mode that abruptly disappears above the transition temperature strongly suggests a first-order nature of the structural phase transition. %\textcolor{red}{The blue shift in P$_{11}$ phonon mode with increasing temperature is inferred as the signature of strong anharmonicity or phonon-phonon scattering. } 
The Fano asymmetry  revealed considerable coupling between phonon modes and electronic degrees of freedom. By First principles phonon calculations along with symmetry analysis, it is found that anharmonicity, minimal strain coupling and electron-phonon interaction are the reason for the first-order phase transition. Our findings offer a comprehensive study of inversion symmetry breaking in a quasi-1D system through phonon mode renormalization and will stimulate further research into the role of symmetry in collective lattice dynamics, including CDW condensates.

\section{Acknowledgements}
This work was supported by the (i) 'Department of Science and Technology, Government of India (Grant No. SRG/2019/000674 and EMR/2016/005437), and (ii) Collaborative Research Scheme (CRS) project proposal(2021/CRS/59/58/760). A.B. thanks CSIR Govt. of India for Research Fellowship with Grant No. 09/080(1109)/2019-EMR-I. M.M.,  A.B., and P.S.R. acknowledge support from Ms. Surabhi Saha for theoretical calculations and inputs. Finally, the authors would like to acknowledge Prof. Tanusri Saha-Dasgupta for the effective discussions and feedback. \\

\textbf{The authors declare no competing financial interests.}\\

\section{Method}

\textbf{(TaSe$_4)_3$I Single Crystal Growth:}
For this study, (TaSe$_4)_3$I single crystals  were grown by the chemical vapor transport (CVT) method. Ta (Alfa Aeser, purity 99.97\%) and Se (Alfa Aeser, purity 99.999\%) powders were combined with iodine pieces (Alfa Aeser, 99.0\%) in a stoichiometric ratio and ground together into a fine powder. The resulting powder was loaded into a vacuum-sealed quartz tube. The tube was then placed in a two-zone furnace for one week, maintained at temperatures of 400$~^\circ$C and 500$~^\circ$C. Hair/fiber-like single crystals were grown along random directions. After cooling to room temperature, the crystals (silver/grey) were collected as a lump of fiber wool by breaking the quartz tube in the air at room temperature ($\approx$ 300 K). The material is stable under ambient conditions. A schematic representation of the CVT method can be found in the supplementary section.

%\subsection{Characterisation}
%\textbf{Crystal structure :} Single crystal XRD (SXRD) characterization at room temperature, as well as other temperatures, was done on a Bruker D8 VENTURE Microfocus diffractometer equipped with PHOTON II Detector, with Mo K$_{\alpha}$ radiation ($\lambda$ = 0.71073 $^{\circ}$). \\
%\textbf{Morphology and elemental analysis :} Crystal Morphology and chemical compositions were confirmed using a high-vacuum scanning electron microscope (SEM)(JEOL LSM-6500). The crystals are formed in ribbon-like fibers of length $\approx$ few mm, width $\approx$ few microns. A field emission scanning electron microscopy (FESEM, Zeiss Sigma) image showing a flat belt-like shape of the crystal is shown in Fig 1 (c). \\
%\textbf{Transport property : } To investigate the electronic properties of the needle-like crystals, we have employed the usual four-probe resistivity
%measurements on a bunch of ribbons as well as on fewer ribbons. absolute Seebeck coefficient measured on the slowly cooled sample on the cooling temperature ramp. \\
%\textbf{Heat capacity measurement :} Heat capacity at $H$ = 0 magnetic field, in the temperature range 100 - 250 K, was measured using the Quantum Design Physical Properties Measurement System (PPMS).\\

\textbf{Morphology:}
To confirm the crystal morphology at room temperature (RT), we employed a high-vacuum scanning electron microscope (SEM) manufactured by JEOL (model LSM-6500). This state-of-the-art equipment allowed us to obtain high-resolution images of the crystals, providing valuable insights into their shape and structure.

\textbf{Electronic transport measurements} were carried out in standard four-probe technique on a few ribbon-like oriented single crystals using a "Keithley 2450 source meter" combined with "Keithley 2182A nanovoltmeter" in Cryogenics 16 Tesla measurement system.

\textbf{Raman spectroscopy measurement: } Raman measurement at room temperature was performed using a 532 nm laser excitation in Horiba T64000 Raman spectrometer (with spot size $\sim$ 1~$\mu$m).

For temperature-dependent Raman spectra, we used a liquid N$_2$ flow cryostat to measure the sample in the range of 85  to 300~K. %This allowed us to probe the crystallographic degrees of freedom in (TaSe$_4$)$_3$I. Additionally, 
we carried out polarized Raman spectroscopy to detect the symmetry-resolved excitations of lattice vibrations. (the schematic is shown in the supplementary section S5 and Fig.~S7). 

\textbf{First-principles calculations} were performed  within the framework of the density functional theory (DFT) as implemented in QUANTUM ESPRESSO \cite{Giannozzi2009} and PHONOPY software \cite{phonopy} using ($1\times 1\times 1$) primitive cell to calculate phonon frequencies at zone-center. The ultra-soft potential is used in DFT to calculate the force constants and passed to PHONOPY to calculate the dynamical matrix.  The distortion configuration of each relevant mode is visualized first and assigned with proper symmetry.

%\textbf{Theoretical calculations:}  

\textbf{Data availability.} The data that support the findings of this study are available from the corresponding author upon
reasonable request.

%\bibliography{RefnTSI}{}


\newpage

\onecolumngrid

\textbf{\Large {\underline {Supplement Materials:} Raman signatures of lattice dynamics across inversion symmetry breaking phase transition in quasi-1D compound, \ch{(TaSe4)3I} }}

\section{Single Crystal growth}
The schematic of the crystal growth process using the chemical vapor transport (CVT) technique is shown here, along with the optical picture of the as-grown (TaSe$_4$)$_3$I (TSI3) crystals.
\begin{figure}[h]
	
	\begin{center}
		\includegraphics[width=1\columnwidth]{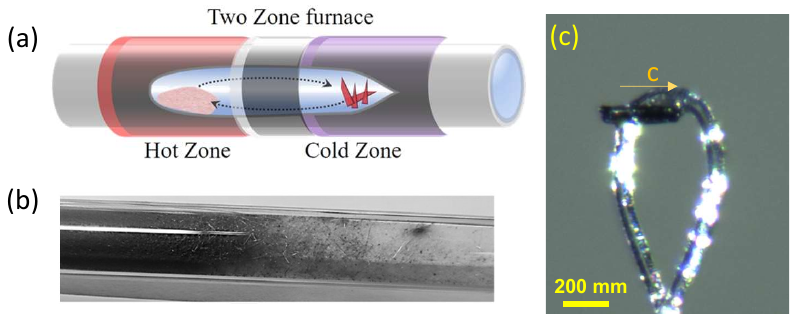}
		\caption {\textbf{Crystal growth and characterization.} (a) Schematic diagram of the synthesis of TSI3 single crystals
			using the CVT method. (b) Picture of an as-grown  sample composed of many needle-like single crystals in a sealed quartz tube. (c) A zoomed-in optical image of a crystal used for single crystal XRD study.}
		\label{CVT}
	\end{center}
\end{figure}
%\section{SEM micrograph}
%The crystals are formed in ribbon-like fibers of length $\approx$ few mm, width $\approx$ few microns. A field emission scanning electron microscopy (FE-SEM, Zeiss Sigma) image showing a flat belt-like shape of the crystal.

\section{Thermal evolution of Raman spectra.}
\begin{figure}
	\begin{center}
		\includegraphics[width=.8\columnwidth]{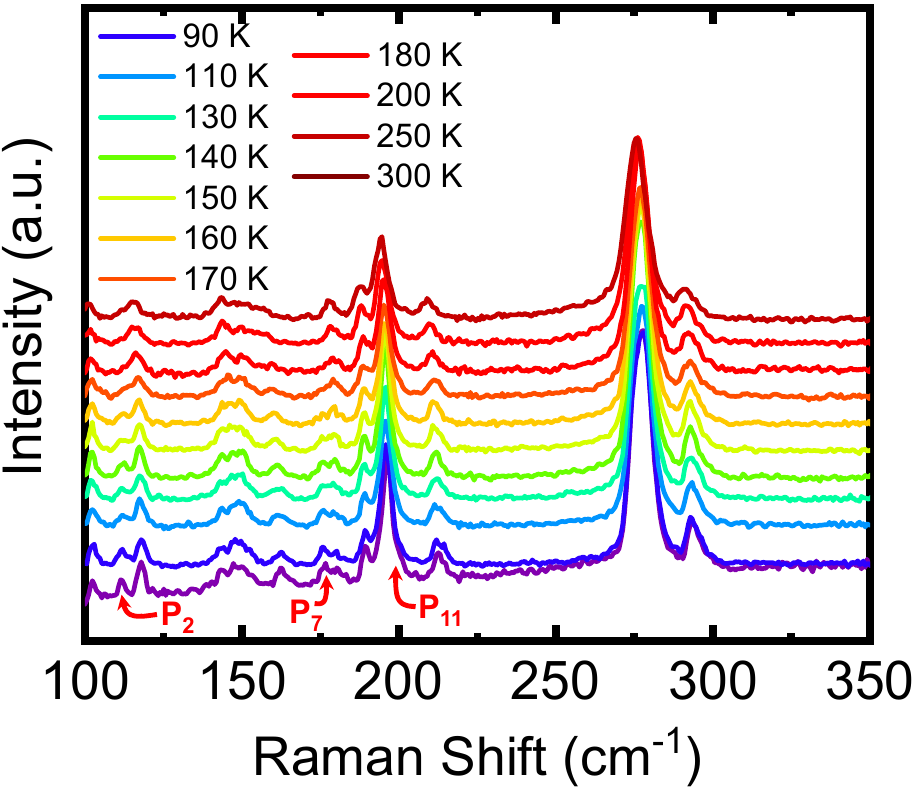}
		\caption {\textbf{Temperature dependent Raman spectra of (TaSe$_4$)$_3$I.}  Temperature-dependent Raman spectra for (TaSe$_4$)$_3$I in the spectral range of 100 - 350 cm$^{-1}$. The sample temperature increases from 90 K (bottom) to 300K (top). The evolution of vibration modes, P$_2$, P$_7$, and P$_{11}$ are also shown here with temperature.}
		\label{Raman-T}
	\end{center}
\end{figure}

\begin{figure}
	\begin{center}
		\includegraphics[width=1\columnwidth]{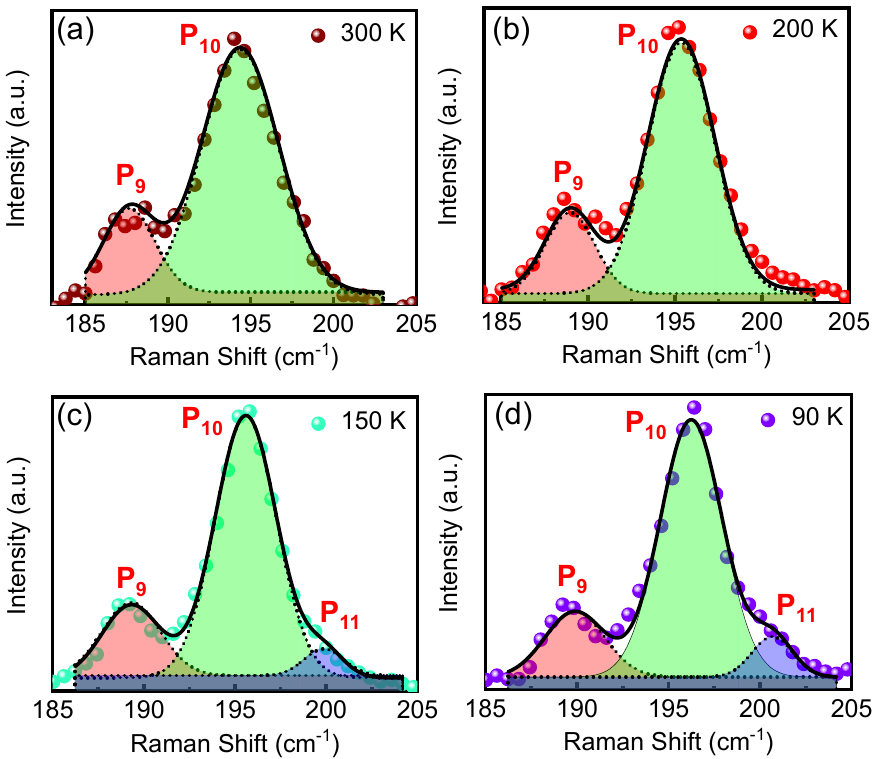}
		\caption {\textbf{Thermal evolution of P$_{11}$ :} Illustrates the appearance of P$_{11}$ mode below 160 K. Scattered solid spheres and black solid lines represent the experimental and resultant simulated curve.}
		\label{P10}
	\end{center}
\end{figure}

To elucidate the nature of the observed phase transition in TSI3, we plot the intensity (a.u.) color-map of the Raman spectra, as a function of temperature(T) and frequency($\omega$) (see Fig.~\ref{Raman-T}). The color variation, ranging from red to purple, represents a decrease in sample temperature, T from 300~K to 90~K. Notably, below 160 K (highlighted in red), the P$_2$, P$_7$, and P$_{11}$ modes become apparent. To provide a clearer visualization of the P$_{11}$ mode, which emerges at lower temperatures, we have presented the thermal evolution of this specific mode for four distinct temperature points (two above and two below the transition) in Fig.~\ref{P10}. This  analysis of the P$_{11}$ mode allows us to examine its behavior and relationship with the phase transition.

\begin{figure}[t!]
	\begin{center}
		\includegraphics[width=1\columnwidth]{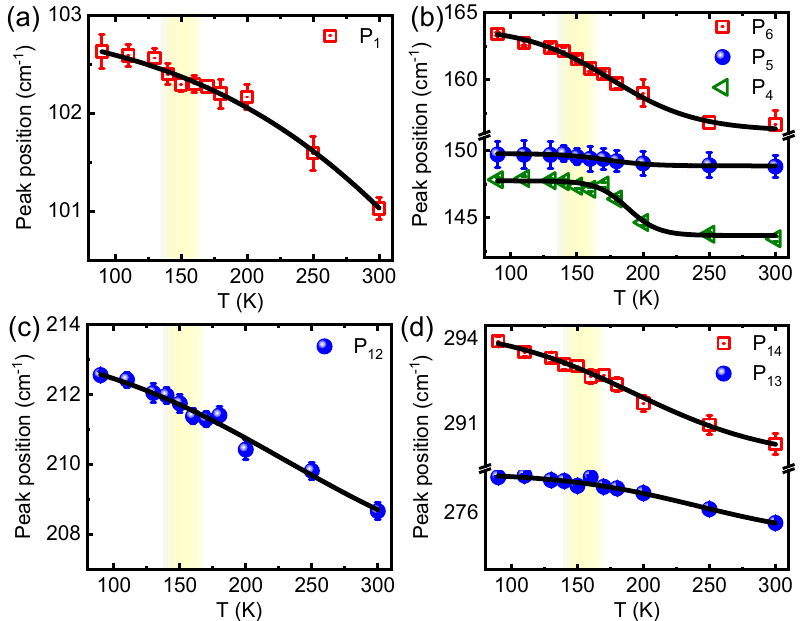}
		\caption {\textbf{Thermal evolution of characteristics frequency of remaining Raman modes.}  Temperature evolution of the self-energy parameters. Scattered red, blue and green symbol are experimental data, whereas the continuous black lines are fitted curves using the symmetric three-phonon coupling model (mentioned in the main text.). The yellow-shaded regions represent the temperature regime of phase transition.}  
	\end{center}
	\label{ModeFreq}
\end{figure}

\begin{figure}
	\begin{center}
		\includegraphics[width=1\columnwidth]{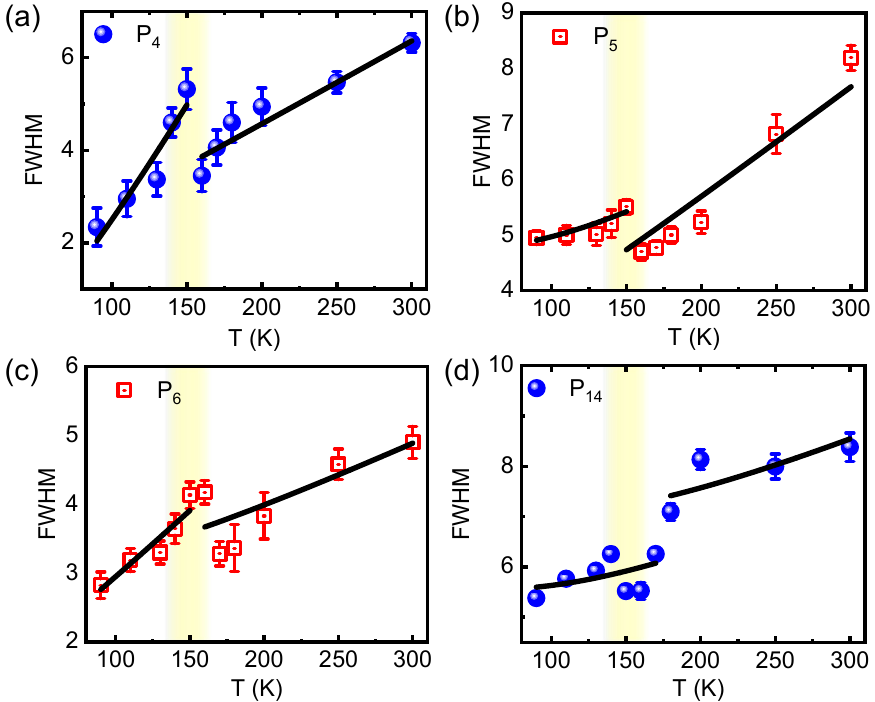}
		\caption {\textbf{Thermal evolution of FWHMs of the Raman modes.} Temperature evolution of the full width at half maxima (FWHM) remaining Raman modes. Scattered red, and blue spheres are experimental data, whereas the continuous black lines are fitted curves according to the symmetric three-phonon coupling model.}  
		\label{S:FWHM}
	\end{center}
\end{figure}

\subsection{Temperature variation of Raman Mode characteristics}
%Temperature evolution of the self-energy parameters is presented in Fig.~S4 and Fig.~\ref{S:FWHM}. The temperature-dependent phonon mode frequency ($\omega$) and linewidth($\Gamma$) are described by symmetric three-phonon coupling models \cite{Menendez1984}, %\begin{equation}
%    \omega(T) = \omega_1 + \frac{\omega^\prime - \omega_1}{1+exp\frac{T-T_0}{dT}}
%\end{equation}

%\begin{equation}
%   \Gamma = \Gamma_0\left(1 + \frac{1}{exp\frac{hc\bar{\nu_1}}{k_BT}-1} + \frac{1}{exp\frac{hc\bar{\nu_2}}{k_BT}-1}\right)
%\end{equation}
%Here, $\omega^\prime$ and $\omega_1$ represent the top and bottom of the sigmoidal curve, respectively \cite{need citation}. $T_0$ is the center point and d$T$ controls the width of the curve. $h$ and $k_B$ are the Planck and Boltzmann constants, respectively. The $c$ is the speed of light and $T$ is the temperature. The $\Gamma_0$ is the asymptotic value of the linewidth at zero temperature. \bar{$\nu_1$} and \bar{$\nu_2$} are two acoustic phonon modes with different wavenumbers with opposite wavevectors.

The temperature evolution of the full width at half maxima (FWHM) is depicted in Fig.~\ref{S:FWHM}. The yellow-shaded regions represent the temperature regime of the phase transition (see Fig.~S4 and Fig.~\ref{S:FWHM}). Their line widths (FWHMs) monotonically increase with increasing temperature, T along with sudden jumps at around $\sim$ T$_S$. More often in solids at high temperatures phonon-phonon interaction increases, resulting in the lowering of phonon lifetime and increase in line widths with increasing temperature. The sudden upturn of FWHM at around T$_S$ strongly suggests an abrupt change in coupling between lattice and phonon degrees of freedom associated with phase transition.

\subsection{Details of Raman modes}

The details of the Raman modes simulated from the theoretical calculation are described below and represented in Fig.6 of main text and Fig.S9 and Fig. S10 of supplementary. principle rotational axis presented in Fig.1 of main text.\\
(i) $A_{1g}$: Non-degenerate 1D mode, that is symmetric with respect to the principle rotational axis, other rotational axis, mirror plane, and inversion. \\
(ii) $A_{1u}$: Non-degenerate 1D mode, which is symmetric with respect to the principle rotational axis (n($C_n$)), other rotational axes ($C_n$)), mirror plane ($m$), and anti-symmetric to the inversion ($\bar{I}$). \\
(iii) $A_{2g}$:  Non-degenerate 1D mode, which is symmetric to the inversion, principle rotation axis,  and anti-symmetric with respect to the rotational axis and mirror plane. \\
(iv) $A_{2u}$:  Non-degenerate 1D mode, which is symmetric to the principle rotational axis and anti-symmetric with respect to the rotation axis ($C_n$), mirror plane, and inversion center. \\
(v) $B_{1g}$:  Non-degenerate 1D mode, which is anti-symmetric with respect to the principle rotational axis and symmetric to mirror plane, inversion center, and other rotational axes~($C_n$). \\
(vi) $B_{1u}$:  Non-degenerate 1D mode, which is anti-symmetric with respect to the principle rotational axis, inversion center, and symmetric to mirror plane (m), another rotational axis~($C_n$). \\
(vii) $B_{2g}$:  Non-degenerate 1D mode, which is anti-symmetric with respect to any rotational axis (n($C_n$) or $C_n$), mirror plane and symmetric to the inversion center ($\bar{I}$). \\
(ix) $B_{2u}$:  Non-degenerate 1D mode, which is anti-symmetric under all symmetry operations (rotation, mirror plane, and inversion center). \\
(x) $E_{u}$:  Double-degenerate 2D mode, which is anti-symmetry to inversion. \\
(xi) $E_{g}$:  Double-degenerate 2D mode, which is symmetric under inversion. 

%\subsection{Connection between Raman shifts with temperature:}
%In the case of layered material, a phenomenological model \cite{PhysRevB.89.224301, Joshi2016,PhysRevB.98.224104} is successfully applied to describe the dependence of shifts of Raman modes with $T$, based on the calculation of M. Balkanski et al. \cite{PhysRevB.28.1928}. The model describes, \\
%\begin{equation}
%\omega(T) = \omega_B + A\times(1+\frac{2}{e^x-1})
%\end{equation}
%here $x$ =  $\frac{\hbar\omega}{k_BT}$, $\omega_B$ = Harmonic phonon frequency at $T$ = 0 K, A = a constant, and an anharmonic term.
% \begin{figure*}[h]
%\includegraphics[width=1\columnwidth]{figs_fano_gaussian.pdf}
%		\caption {\textbf{Quantitive analysis of the P$_6$ mode at different temperatures.} (a)-(c) The blue solid lines represent the line shape obtained as the convolution of the Fano profile and Gaussian, whereas the green solid lines represent Lorentz profiles.}  
%  \label{S:Fano}
%\end{figure*}
%\section{Fano Fitting}

%The mode P$_{6}$, exhibit significant asymmetric line shapes and fitted with the Fano-Gaussian convolution as discussed also in the main text. The quantative analysis of this peaks is performed using both the symmetric Lorentzian fit and the Fano-Lorentzian convolution. The comparison between the model fit and the experimental data for P$_{6}$ mode is presented in the Fig. \ref{S:Fano} (a-c).

\section{Polarisation dependent Raman Spectroscopy}
%To perform a polarization-dependent Raman spectroscopy study, we did a group theory analysis of both phases and find out the selection rule. These selection rules are shown in table II. We can find a notation that expresses the orientation of the crystal with respect to the polarization of the laser in both the excitation and analyzing directions. This notation is called Porto's notation, and consists of four-letter: A(BC)D\\
%Where,\\
%A,D- directions of the propagation of incident ($k_i$) and scattered ($k_s$) light,\\
%B,c- directions of the polarization incident ($E_i$) and scattered ($E_s$) light.\\

\subsection{Measurement setup}
\begin{figure*}[h]
	\includegraphics[width=1\columnwidth]{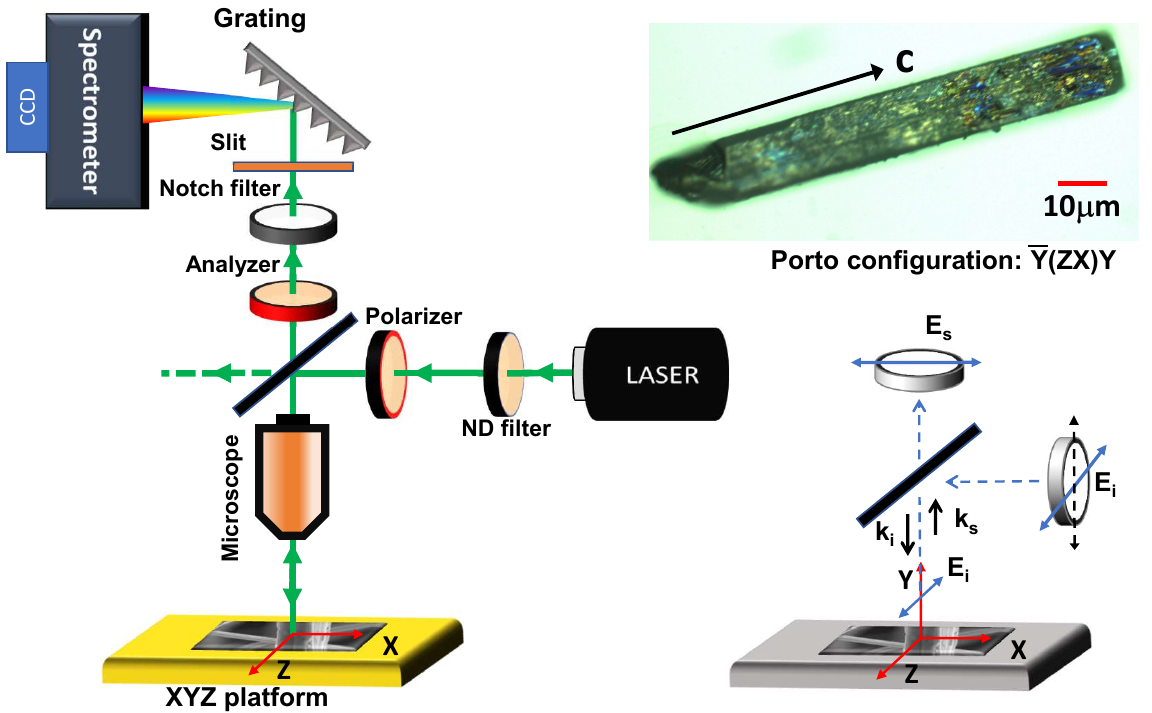}
	\caption {\textbf{Experimental setup :} Schematic of the Raman Spectroscopy measurement setup.} 
	\label{experimental setup}
\end{figure*}

The setup for angle-resolved Raman spectroscopy is depicted in Fig.\ref{experimental setup} at the beginning the linearly polarised light touches the sample, thereafter the scattered light is passed through a polariser oriented either parallel or perpendicular to the polarisation of the incident beam.
\subsection{Porto notation}
The orientation of the crystal w.r.t the polarization of the laser in both the excitation and analyzing directions; scattering geometry is expressed in this notation, termed as Porto notation (P.N). This notation was first coined by S.P.S. Porto in 1966 \cite{PhysRev.142.570}. The general representation of Porto notation is, 
\begin{equation}
	P.N = A(BC)D,   
\end{equation}
A = The direction of the propagation of the incident light.(k$_i$)\\
B = The direction of the polarization of the incident light. (E$_i$)\\
C = The direction of the polarization of the scattered light. (E$_s$)\\
D = The direction of the propagation of the scattered light. (k$_s$)\\
Example: where $\overline{a}(bb)c$ implies the photons are incident in the $\overline{a}$ direction and directly back-scattered light is detected along $c$ direction. Moreover, the direction of the polarization of the incident and scattered light is along the $b$ direction.

\begin{figure*}[b!]
	\includegraphics[width=1\columnwidth]{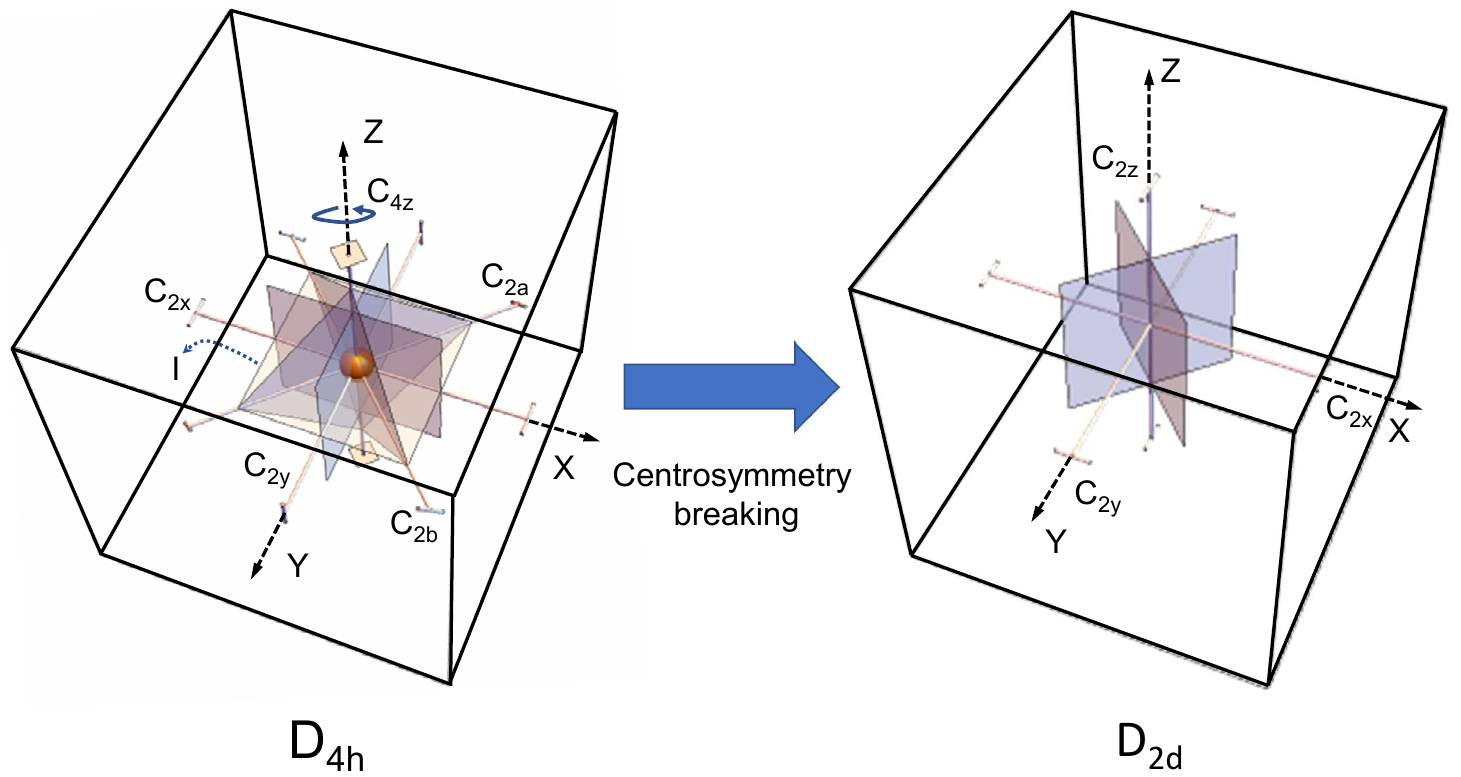}
	\caption {Point group symmetry actions of RT(D$_{4h}$) and LT(D$_{2d}$) phase (generated by GTPack~\cite{gtpack1}).}  
	\label{groupTheory}
\end{figure*}

\section{Group analysis and Landau free energy functional of the phase transition:}

Group elements of D$_{4h}$:
$\left\{\text{Ee},\overset{\text{}}{\text{C}}_{\text{2z}}^{\text{}},\overset{\text{}}{\text{C}}_{\text{2y}}^{\text{}},\overset{\text{}}{\text{C}}_{\text{2b}}^{\text{}},\overset{\text{}}{\text{C}}_{\text{2a}}^{\text{}},\overset{\text{}}{\text{C}}_{\text{2x}}^{\text{}},\overset{\text{}}{\text{C}}_{\text{4z}}^{-1},\overset{\text{}}{\text{C}}_{\text{4z}}^{\text{}},\text{IEe},\overset{\text{}}{\text{IC}}_{\text{2x}}^{\text{}},\overset{\text{}}{\text{IC}}_{\text{2a}}^{\text{}},\overset{\text{}}{\text{IC}}_{\text{2b}}^{\text{}},\overset{\text{}}{\text{IC}}_{\text{2y}}^{\text{}},\overset{\text{}}{\text{IC}}_{\text{2z}}^{\text{}},\overset{\text{}}{\text{IC}}_{\text{4z}}^{\text{}},\overset{\text{}}{\text{IC}}_{\text{4z}}^{-1}\right\}$

Group elements of D$_{2d}$: $\left\{\text{Ee},\overset{\text{}}{\text{C}}_{\text{2z}}^{\text{}},\overset{\text{}}{\text{C}}_{\text{2y}}^{\text{}},\overset{\text{}}{\text{C}}_{\text{2x}}^{\text{}},\overset{\text{}}{\text{IC}}_{\text{2c}}^{\text{}},\overset{\text{}}{\text{IC}}_{\text{2d}}^{\text{}},\overset{\text{}}{\text{IC}}_{\text{4y}}^{-1},\overset{\text{}}{\text{IC}}_{\text{4y}}^{\text{}}\right\}$

Character table of  D$_{4h}$ point group (point group of RT symmetric phase)\cite{gtpack1,gtpack2}: \\

\begin{center}
\[	\begin{array}{c|cccccccccc}
		\text{} & \text{Ee} & 2 \overset{\text{}}{\text{C}}_{\text{2y}}^{\text{}} & 2 \overset{\text{}}{\text{IC}}_{\text{2x}}^{\text{}} & 2 \overset{\text{}}{\text{C}}_{\text{2b}}^{\text{}} & 2 \overset{\text{}}{\text{IC}}_{\text{2a}}^{\text{}} & 2 \overset{\text{}}{\text{IC}}_{\text{4z}}^{\text{}} & 2 \overset{\text{}}{\text{C}}_{\text{4z}}^{-1} & \text{IEe} & \overset{\text{}}{\text{C}}_{\text{2z}}^{\text{}} & \overset{\text{}}{\text{IC}}_{\text{2z}}^{\text{}} \\
		\hline
		\Gamma ^1 & 1 & 1 & 1 & 1 & 1 & 1 & 1 & 1 & 1 & 1 \\
		\Gamma ^2 & 1 & -1 & -1 & -1 & -1 & 1 & 1 & 1 & 1 & 1 \\
		\Gamma ^3 & 1 & -1 & -1 & 1 & 1 & -1 & -1 & 1 & 1 & 1 \\
		\Gamma ^4 & 1 & -1 & 1 & -1 & 1 & -1 & 1 & -1 & 1 & -1 \\
		\Gamma ^5 & 1 & -1 & 1 & 1 & -1 & 1 & -1 & -1 & 1 & -1 \\
		\Gamma ^6 & 1 & 1 & -1 & -1 & 1 & 1 & -1 & -1 & 1 & -1 \\
		\Gamma ^7 & 1 & 1 & -1 & 1 & -1 & -1 & 1 & -1 & 1 & -1 \\
		\Gamma ^8 & 1 & 1 & 1 & -1 & -1 & -1 & -1 & 1 & 1 & 1 \\
		\Gamma ^9 & 2 & 0 & 0 & 0 & 0 & 0 & 0 & -2 & -2 & 2 \\
		\Gamma ^{10} & 2 & 0 & 0 & 0 & 0 & 0 & 0 & 2 & -2 & -2 \\
	\end{array} \]
\end{center}

\newpage
Bethe to Mulliken( used in GTPack\cite{gtpack2} )  to Miller and Love( used in ISOTROPY suite\cite{isotropy_suite}) notation of irredicible representations( irreps):
\begin{center}
	$\Gamma ^1 \rightarrow  \text{A}_{\text{1g}}  \rightarrow \Gamma^+_1 $\\
	$\Gamma ^2 \rightarrow \text{A}_{\text{2g}} \rightarrow \Gamma^+_3$\\
	$ \Gamma ^3 \rightarrow \text{B}_{\text{2g}}  \rightarrow \Gamma^+_4$ \\
	$ \Gamma ^4 \rightarrow \text{A}_{\text{2u}} \rightarrow \Gamma^-_3 $\\
	$ \Gamma ^5 \rightarrow \text{B}_{\text{2u}}  \rightarrow \Gamma^-_4 $\\
	$ \Gamma ^6 \rightarrow  \text{B}_{\text{1u}}  \rightarrow \Gamma^-_2$ \\
	$ \Gamma ^7 \rightarrow \text{A}_{\text{1u}} \rightarrow \Gamma^-_1$\\
	$ \Gamma ^8 \rightarrow \text{B}_{\text{1g}}  \rightarrow \Gamma^+_2 $\\
	$ \Gamma ^9  \rightarrow \text{E}_{\text{u}}  \rightarrow \Gamma^-_5$\\
	$\Gamma ^{10} \rightarrow \text{E}_{\text{g}}  \rightarrow \Gamma^+_5$ \\ 
\end{center}

Checking the invarance of monomial terms, under the action of above irreps, in expansion of $(x+y+z)^{n}$ upto n=6.\\
n=1:\\
\begin{center}
\[	\begin{array}{c|cccccccccc}
		& \Gamma ^1 & \Gamma ^2 & \Gamma ^3 & \Gamma ^4 & \Gamma ^5 & \Gamma ^6 & \Gamma ^7 & \Gamma ^8 & \Gamma ^9 & \Gamma ^{10} \\
		\hline
		x & 0 & 0 & 0 & 0 & 0 & 0 & 0 & 0 & x & 0 \\
		y & 0 & 0 & 0 & 0 & 0 & 0 & 0 & 0 & y & 0 \\
		z & 0 & 0 & 0 & z & 0 & 0 & 0 & 0 & 0 & 0 \\
	\end{array}\]
\end{center}

\newpage
n=2:\\
\begin{center}
\[	\begin{array}{c|cccccccccc}
		& \Gamma ^1 & \Gamma ^2 & \Gamma ^3 & \Gamma ^4 & \Gamma ^5 & \Gamma ^6 & \Gamma ^7 & \Gamma ^8 & \Gamma ^9 & \Gamma ^{10} \\
		\hline
		x^2 & \frac{1}{2} \left(x^2+y^2\right) & 0 & 0 & 0 & 0 & 0 & 0 & \frac{1}{2} \left(x^2-y^2\right) & 0 & 0 \\
		2 x y & 0 & 0 & 2 x y & 0 & 0 & 0 & 0 & 0 & 0 & 0 \\
		2 x z & 0 & 0 & 0 & 0 & 0 & 0 & 0 & 0 & 0 & 2 x z \\
		y^2 & \frac{1}{2} \left(x^2+y^2\right) & 0 & 0 & 0 & 0 & 0 & 0 & \frac{1}{2} \left(y^2-x^2\right) & 0 & 0 \\
		2 y z & 0 & 0 & 0 & 0 & 0 & 0 & 0 & 0 & 0 & 2 y z \\
		z^2 & z^2 & 0 & 0 & 0 & 0 & 0 & 0 & 0 & 0 & 0 \\
	\end{array} \]
\end{center}

%\newpage
n=3:\\
\begin{center}
\[	\begin{array}{c|cccccccccc}
		& \Gamma ^1 & \Gamma ^2 & \Gamma ^3 & \Gamma ^4 & \Gamma ^5 & \Gamma ^6 & \Gamma ^7 & \Gamma ^8 & \Gamma ^9 & \Gamma ^{10} \\
		\hline
		x^3 & 0 & 0 & 0 & 0 & 0 & 0 & 0 & 0 & x^3 & 0 \\
		3 x^2 y & 0 & 0 & 0 & 0 & 0 & 0 & 0 & 0 & 3 x^2 y & 0 \\
		3 x^2 z & 0 & 0 & 0 & \frac{3}{2} z \left(x^2+y^2\right) & \frac{3}{2} z \left(x^2-y^2\right) & 0 & 0 & 0 & 0 & 0 \\
		3 x y^2 & 0 & 0 & 0 & 0 & 0 & 0 & 0 & 0 & 3 x y^2 & 0 \\
		6 x y z & 0 & 0 & 0 & 0 & 0 & 6 x y z & 0 & 0 & 0 & 0 \\
		3 x z^2 & 0 & 0 & 0 & 0 & 0 & 0 & 0 & 0 & 3 x z^2 & 0 \\
		y^3 & 0 & 0 & 0 & 0 & 0 & 0 & 0 & 0 & y^3 & 0 \\
		3 y^2 z & 0 & 0 & 0 & \frac{3}{2} z \left(x^2+y^2\right) & \frac{1}{2} (-3) z \left(x^2-y^2\right) & 0 & 0 & 0 & 0 & 0 \\
		3 y z^2 & 0 & 0 & 0 & 0 & 0 & 0 & 0 & 0 & 3 y z^2 & 0 \\
		z^3 & 0 & 0 & 0 & z^3 & 0 & 0 & 0 & 0 & 0 & 0 \\
	\end{array}\]
\end{center}

%\newpage
n=4:\\
\begin{center}
\[	\begin{array}{c|cccccccccc}
		& \Gamma ^1 & \Gamma ^2 & \Gamma ^3 & \Gamma ^4 & \Gamma ^5 & \Gamma ^6 & \Gamma ^7 & \Gamma ^8 & \Gamma ^9 & \Gamma ^{10} \\
		\hline
		x^4 & \frac{1}{2} \left(x^4+y^4\right) & 0 & 0 & 0 & 0 & 0 & 0 & \frac{1}{2} \left(x^4-y^4\right) & 0 & 0 \\
		4 x^3 y & 0 & 2 x y \left(x^2-y^2\right) & 2 x y \left(x^2+y^2\right) & 0 & 0 & 0 & 0 & 0 & 0 & 0 \\
		4 x^3 z & 0 & 0 & 0 & 0 & 0 & 0 & 0 & 0 & 0 & 4 x^3 z \\
		6 x^2 y^2 & 6 x^2 y^2 & 0 & 0 & 0 & 0 & 0 & 0 & 0 & 0 & 0 \\
		12 x^2 y z & 0 & 0 & 0 & 0 & 0 & 0 & 0 & 0 & 0 & 12 x^2 y z \\
		6 x^2 z^2 & 3 z^2 \left(x^2+y^2\right) & 0 & 0 & 0 & 0 & 0 & 0 & 3 z^2 \left(x^2-y^2\right) & 0 & 0 \\
		4 x y^3 & 0 & 2 x y^3-2 x^3 y & 2 x y \left(x^2+y^2\right) & 0 & 0 & 0 & 0 & 0 & 0 & 0 \\
		12 x y^2 z & 0 & 0 & 0 & 0 & 0 & 0 & 0 & 0 & 0 & 12 x y^2 z \\
		12 x y z^2 & 0 & 0 & 12 x y z^2 & 0 & 0 & 0 & 0 & 0 & 0 & 0 \\
		4 x z^3 & 0 & 0 & 0 & 0 & 0 & 0 & 0 & 0 & 0 & 4 x z^3 \\
		y^4 & \frac{1}{2} \left(x^4+y^4\right) & 0 & 0 & 0 & 0 & 0 & 0 & \frac{1}{2} \left(y^4-x^4\right) & 0 & 0 \\
		4 y^3 z & 0 & 0 & 0 & 0 & 0 & 0 & 0 & 0 & 0 & 4 y^3 z \\
		6 y^2 z^2 & 3 z^2 \left(x^2+y^2\right) & 0 & 0 & 0 & 0 & 0 & 0 & 3 z^2 \left(y^2-x^2\right) & 0 & 0 \\
		4 y z^3 & 0 & 0 & 0 & 0 & 0 & 0 & 0 & 0 & 0 & 4 y z^3 \\
		z^4 & z^4 & 0 & 0 & 0 & 0 & 0 & 0 & 0 & 0 & 0 \\
	\end{array}\]
\end{center}

\newpage
n=5:
\begin{center}
 \[ \begin{array}{c|cccccccccc}
			& \Gamma ^1 & \Gamma ^2 & \Gamma ^3 & \Gamma ^4 & \Gamma ^5 & \Gamma ^6 & \Gamma ^7 & \Gamma ^8 & \Gamma ^9 & \Gamma ^{10} \\
			\hline
			x^5 & 0 & 0 & 0 & 0 & 0 & 0 & 0 & 0 & x^5 & 0 \\
			5 x^4 y & 0 & 0 & 0 & 0 & 0 & 0 & 0 & 0 & 5 x^4 y & 0 \\
			5 x^4 z & 0 & 0 & 0 & \frac{5}{2} z \left(x^4+y^4\right) & \frac{5}{2} z \left(x^4-y^4\right) & 0 & 0 & 0 & 0 & 0 \\
			10 x^3 y^2 & 0 & 0 & 0 & 0 & 0 & 0 & 0 & 0 & 10 x^3 y^2 & 0 \\
			20 x^3 y z & 0 & 0 & 0 & 0 & 0 & 10 x y z \left(x^2+y^2\right) & 10 x y z \left(x^2-y^2\right) & 0 & 0 & 0 \\
			10 x^3 z^2 & 0 & 0 & 0 & 0 & 0 & 0 & 0 & 0 & 10 x^3 z^2 & 0 \\
			10 x^2 y^3 & 0 & 0 & 0 & 0 & 0 & 0 & 0 & 0 & 10 x^2 y^3 & 0 \\
			30 x^2 y^2 z & 0 & 0 & 0 & 30 x^2 y^2 z & 0 & 0 & 0 & 0 & 0 & 0 \\
			30 x^2 y z^2 & 0 & 0 & 0 & 0 & 0 & 0 & 0 & 0 & 30 x^2 y z^2 & 0 \\
			10 x^2 z^3 & 0 & 0 & 0 & 5 z^3 \left(x^2+y^2\right) & 5 z^3 \left(x^2-y^2\right) & 0 & 0 & 0 & 0 & 0 \\
			5 x y^4 & 0 & 0 & 0 & 0 & 0 & 0 & 0 & 0 & 5 x y^4 & 0 \\
			20 x y^3 z & 0 & 0 & 0 & 0 & 0 & 10 x y z \left(x^2+y^2\right) & 10 x y z \left(y^2-x^2\right) & 0 & 0 & 0 \\
			30 x y^2 z^2 & 0 & 0 & 0 & 0 & 0 & 0 & 0 & 0 & 30 x y^2 z^2 & 0 \\
			20 x y z^3 & 0 & 0 & 0 & 0 & 0 & 20 x y z^3 & 0 & 0 & 0 & 0 \\
			5 x z^4 & 0 & 0 & 0 & 0 & 0 & 0 & 0 & 0 & 5 x z^4 & 0 \\
			y^5 & 0 & 0 & 0 & 0 & 0 & 0 & 0 & 0 & y^5 & 0 \\
			5 y^4 z & 0 & 0 & 0 & \frac{5}{2} z \left(x^4+y^4\right) & \frac{1}{2} (-5) z \left(x^4-y^4\right) & 0 & 0 & 0 & 0 & 0 \\
			10 y^3 z^2 & 0 & 0 & 0 & 0 & 0 & 0 & 0 & 0 & 10 y^3 z^2 & 0 \\
			10 y^2 z^3 & 0 & 0 & 0 & 5 z^3 \left(x^2+y^2\right) & 5 z^3 \left(y^2-x^2\right) & 0 & 0 & 0 & 0 & 0 \\
			5 y z^4 & 0 & 0 & 0 & 0 & 0 & 0 & 0 & 0 & 5 y z^4 & 0 \\
			z^5 & 0 & 0 & 0 & z^5 & 0 & 0 & 0 & 0 & 0 & 0 
	       \end{array} \] 
\end{center}

%\newpage
n=6
\setlength{\tabcolsep}{0pt} 
\renewcommand{\arraystretch}{0.0} 
\begin{center}
\[\begin{array}{c|cccccccccc}
			& \Gamma ^1 & \Gamma ^2 & \Gamma ^3 & \Gamma ^4 & \Gamma ^5 & \Gamma ^6 & \Gamma ^7 & \Gamma ^8 & \Gamma ^9 & \Gamma ^{10} \\
			\hline
			x^6 & \frac{1}{2} \left(x^6+y^6\right) & 0 & 0 & 0 & 0 & 0 & 0 & \frac{1}{2} \left(x^6-y^6\right) & 0 & 0 \\
			6 x^5 y & 0 & 3 x y \left(x^4-y^4\right) & 3 x y \left(x^4+y^4\right) & 0 & 0 & 0 & 0 & 0 & 0 & 0 \\
			6 x^5 z & 0 & 0 & 0 & 0 & 0 & 0 & 0 & 0 & 0 & 6 x^5 z \\
			15 x^4 y^2 & \frac{15}{2} x^2 y^2 \left(x^2+y^2\right) & 0 & 0 & 0 & 0 & 0 & 0 & \frac{15}{2} x^2 y^2 \left(x^2-y^2\right) & 0 & 0 \\
			30 x^4 y z & 0 & 0 & 0 & 0 & 0 & 0 & 0 & 0 & 0 & 30 x^4 y z \\
			15 x^4 z^2 & \frac{15}{2} z^2 \left(x^4+y^4\right) & 0 & 0 & 0 & 0 & 0 & 0 & \frac{15}{2} z^2 \left(x^4-y^4\right) & 0 & 0 \\
			20 x^3 y^3 & 0 & 0 & 20 x^3 y^3 & 0 & 0 & 0 & 0 & 0 & 0 & 0 \\
			60 x^3 y^2 z & 0 & 0 & 0 & 0 & 0 & 0 & 0 & 0 & 0 & 60 x^3 y^2 z \\
			60 x^3 y z^2 & 0 & 30 x y z^2 \left(x^2-y^2\right) & 30 x y z^2 \left(x^2+y^2\right) & 0 & 0 & 0 & 0 & 0 & 0 & 0 \\
			20 x^3 z^3 & 0 & 0 & 0 & 0 & 0 & 0 & 0 & 0 & 0 & 20 x^3 z^3 \\
			15 x^2 y^4 & \frac{15}{2} x^2 y^2 \left(x^2+y^2\right) & 0 & 0 & 0 & 0 & 0 & 0 & \frac{1}{2} (-15) x^2 y^2 \left(x^2-y^2\right) & 0 & 0 \\
			60 x^2 y^3 z & 0 & 0 & 0 & 0 & 0 & 0 & 0 & 0 & 0 & 60 x^2 y^3 z \\
			90 x^2 y^2 z^2 & 90 x^2 y^2 z^2 & 0 & 0 & 0 & 0 & 0 & 0 & 0 & 0 & 0 \\
			60 x^2 y z^3 & 0 & 0 & 0 & 0 & 0 & 0 & 0 & 0 & 0 & 60 x^2 y z^3 \\
			15 x^2 z^4 & \frac{15}{2} z^4 \left(x^2+y^2\right) & 0 & 0 & 0 & 0 & 0 & 0 & \frac{15}{2} z^4 \left(x^2-y^2\right) & 0 & 0 \\
			6 x y^5 & 0 & 3 x y^5-3 x^5 y & 3 x y \left(x^4+y^4\right) & 0 & 0 & 0 & 0 & 0 & 0 & 0 \\
			30 x y^4 z & 0 & 0 & 0 & 0 & 0 & 0 & 0 & 0 & 0 & 30 x y^4 z \\
			60 x y^3 z^2 & 0 & 30 x y z^2 \left(y^2-x^2\right) & 30 x y z^2 \left(x^2+y^2\right) & 0 & 0 & 0 & 0 & 0 & 0 & 0 \\
			60 x y^2 z^3 & 0 & 0 & 0 & 0 & 0 & 0 & 0 & 0 & 0 & 60 x y^2 z^3 \\
			30 x y z^4 & 0 & 0 & 30 x y z^4 & 0 & 0 & 0 & 0 & 0 & 0 & 0 \\
			6 x z^5 & 0 & 0 & 0 & 0 & 0 & 0 & 0 & 0 & 0 & 6 x z^5 \\
			y^6 & \frac{1}{2} \left(x^6+y^6\right) & 0 & 0 & 0 & 0 & 0 & 0 & \frac{1}{2} \left(y^6-x^6\right) & 0 & 0 \\
			6 y^5 z & 0 & 0 & 0 & 0 & 0 & 0 & 0 & 0 & 0 & 6 y^5 z \\
			15 y^4 z^2 & \frac{15}{2} z^2 \left(x^4+y^4\right) & 0 & 0 & 0 & 0 & 0 & 0 & \frac{1}{2} (-15) z^2 \left(x^4-y^4\right) & 0 & 0 \\
			20 y^3 z^3 & 0 & 0 & 0 & 0 & 0 & 0 & 0 & 0 & 0 & 20 y^3 z^3 \\
			15 y^2 z^4 & \frac{15}{2} z^4 \left(x^2+y^2\right) & 0 & 0 & 0 & 0 & 0 & 0 & \frac{1}{2} (-15) z^4 \left(x^2-y^2\right) & 0 & 0 \\
			6 y z^5 & 0 & 0 & 0 & 0 & 0 & 0 & 0 & 0 & 0 & 6 y z^5 \\
			z^6 & z^6 & 0 & 0 & 0 & 0 & 0 & 0 & 0 & 0 & 0 \\
	\end{array}\]
\end{center}

From ISOTROPY\cite{isotropy_suite,group-subg,group-subg_phase} subgroup calculation, $\Gamma^1$, and $\Gamma^6$ modes are related to order parameters(strain/displacement), and only $\Gamma^6$ mode causes continuous phase transition. Although $\Gamma^6$ mode or B$_{1u}$ mode (h  ere P$_{10}$ peak which can be allowed both in RT as  B$_{1u}$ mode and LT as  B$_1$ mode by symmetry) appears to be the cause of phase transition, macroscopic strain from $\Gamma^1$ mode plays a crucial role here due to its symmetry( strain tensor elements associated to irrep $\Gamma_1$ are $\epsilon_{zz}$ and $\epsilon_{xx}+\epsilon_{yy}$).
Continuous phase transition demands gradual softening of the order parameter to zero, not sudden disappearance( in the present case, P$_{10}$ peak disappears with increasing temperature). Two possible order parameters, one for strain ($\Gamma^1$) and another  displacive  in nature($\Gamma^6$), can both be present with minimal coupling. 
Keeping the experimental data in mind, we can have a general landau function considering the displacement of the Ta(3) atom along the Ta-Ta chain, as our order parameter, with elastic strain\cite{ Krumhansl1992, Krumhansl1990Aug, Krumhansl1989Feb, m.t.dove}.
\begin{figure*}[b]
	\includegraphics[width=0.7\columnwidth]{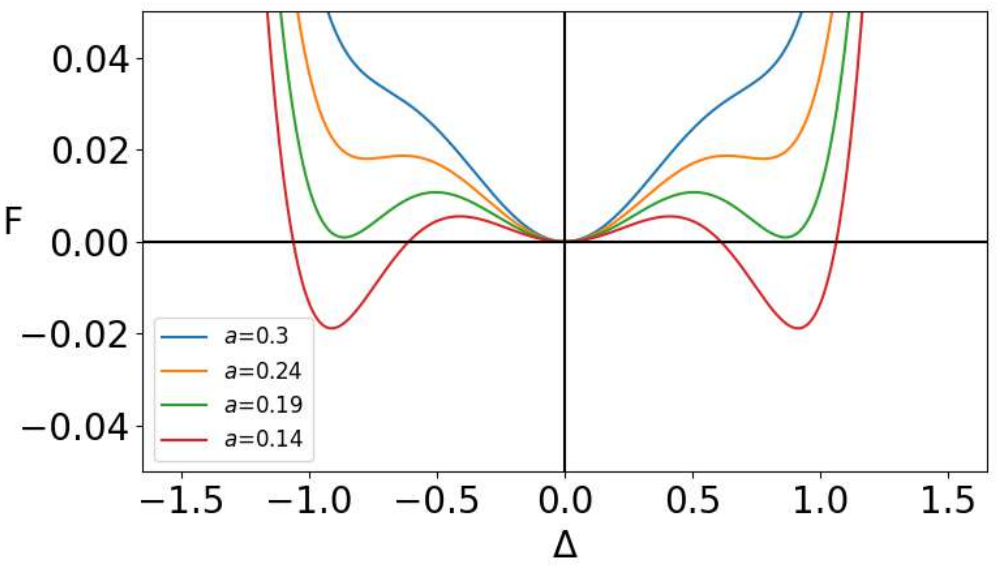}
	\caption {Free energy vs order-parameter with parameter $a$ in eqn(4)} 
	\label{FreeEnergy}
\end{figure*}

$F=f_{\text{elastic}}+ f_{\text{dynamic}}$\\
$f_{\text{elastic}}=\beta ed^2+\frac{1}{2}\alpha  e^2$\\
$f_{\text{dynamic}}=\frac{1 }{2}\lambda d^2+\frac{1}{4}\delta  d^4+\frac{1}{6}\gamma  d^6$\\
where elastic strain $e$ is coupled to displacement $d$. As strain is in equilibrium with displacement\\
$(\frac{\partial F}{\partial e})_d=0$\\
$e=-\frac{\beta d^2}{\alpha}$\\
putting this value of $e$ in the expression of free energy F\\
$F=\frac{1}{2}\lambda d^2+\frac{1}{4} \left(\delta -\frac{\alpha  \beta ^2}{2}\right)d^4+\frac{1}{6}\gamma d^6$\\
We want to scale the order parameter so that the free energy can be cast into a single parameter functional of the order parameter.\\
$b=\delta -\frac{\alpha  \beta ^2}{2}$\\
$d_0^2=\frac{b}{\gamma }$\\
$\Delta =\frac{d}{d_0}$ and $a=\frac{\lambda \gamma }{b^2},$
\begin{equation}
	\begin{split}
		F'=\frac{a}{2}\Delta^2+sgn\ b \ \frac{1}{4}\Delta ^4+\frac{1}{6}\Delta ^6   
	\end{split}
\end{equation}
If $sgn\ b=-1$ and $\gamma> 0$ the transition is of first order in nature, if $sgn\ b=+1$ and $\lambda>0$
it is a second-order or continuous one. The strength of the strain coupling parameter ($\beta$) and overall strain($\alpha$) is high enough to surpass the dynamic anharmonic phonon-phonon scattering strength($\delta$). Therefore anharmonicity and strain coupling are both needed to have such a first-order phase transition. Such first principle calculation is done For ferroelectric phase transition PbTiO$_3$\cite{Waghmare1997}, where in BaTiO$_3$, it is continuous phase transition\cite{batio3}.

\section{Vibration Symmetry Representations of Raman modes}
Here we present Calculated vibration spectra for all the observed corresponding to Raman-active modes (except P$_2$, P$_6$, P$_{11}$) at RT and LT.

\begin{figure*}[b]
	\includegraphics[width=1\columnwidth]{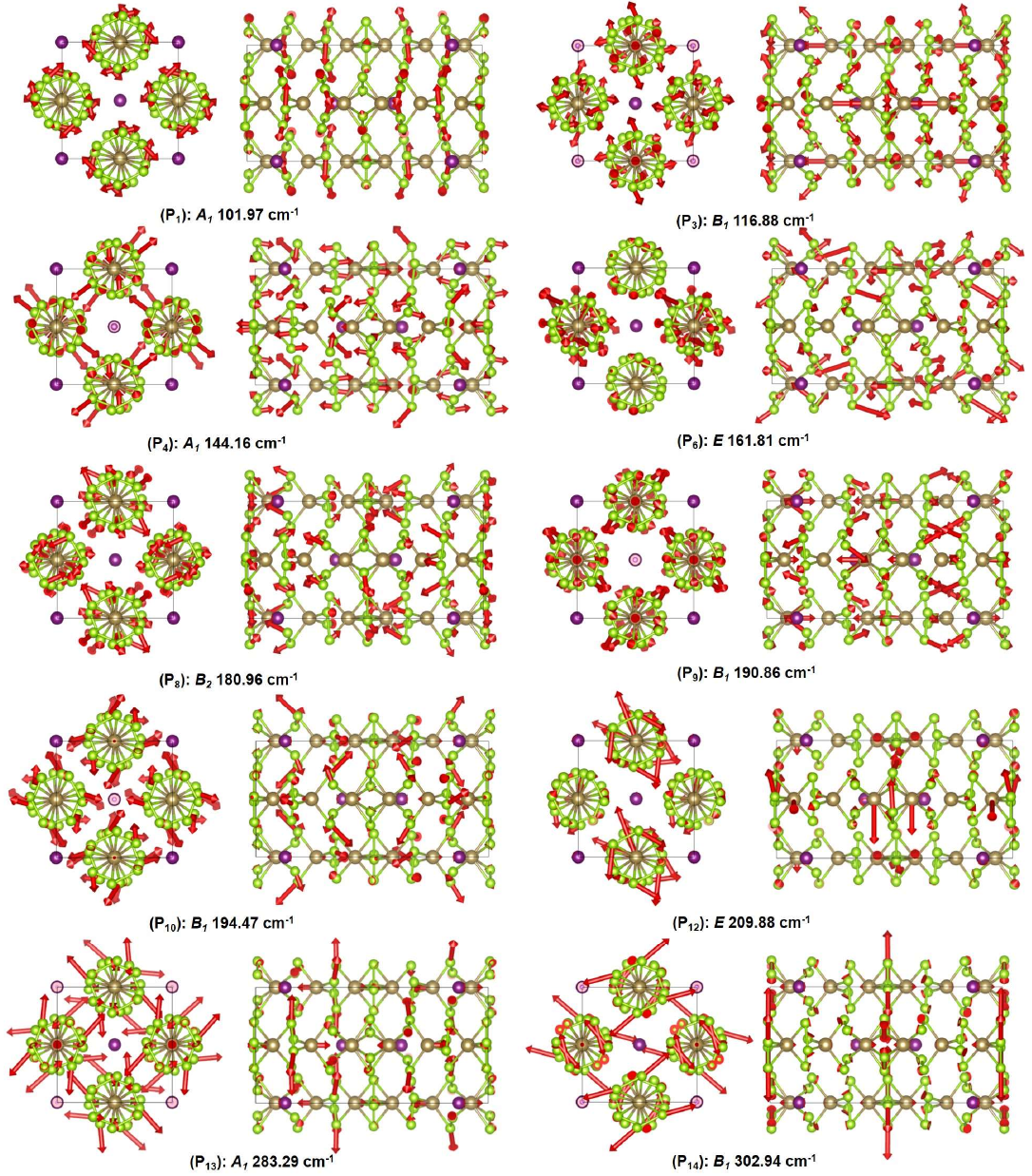}\caption {\textbf{Vibration symmetry representation at low temperature\cite{vesta}}}  
\end{figure*}

\begin{figure*}[h!]
	\includegraphics[width=1\columnwidth]{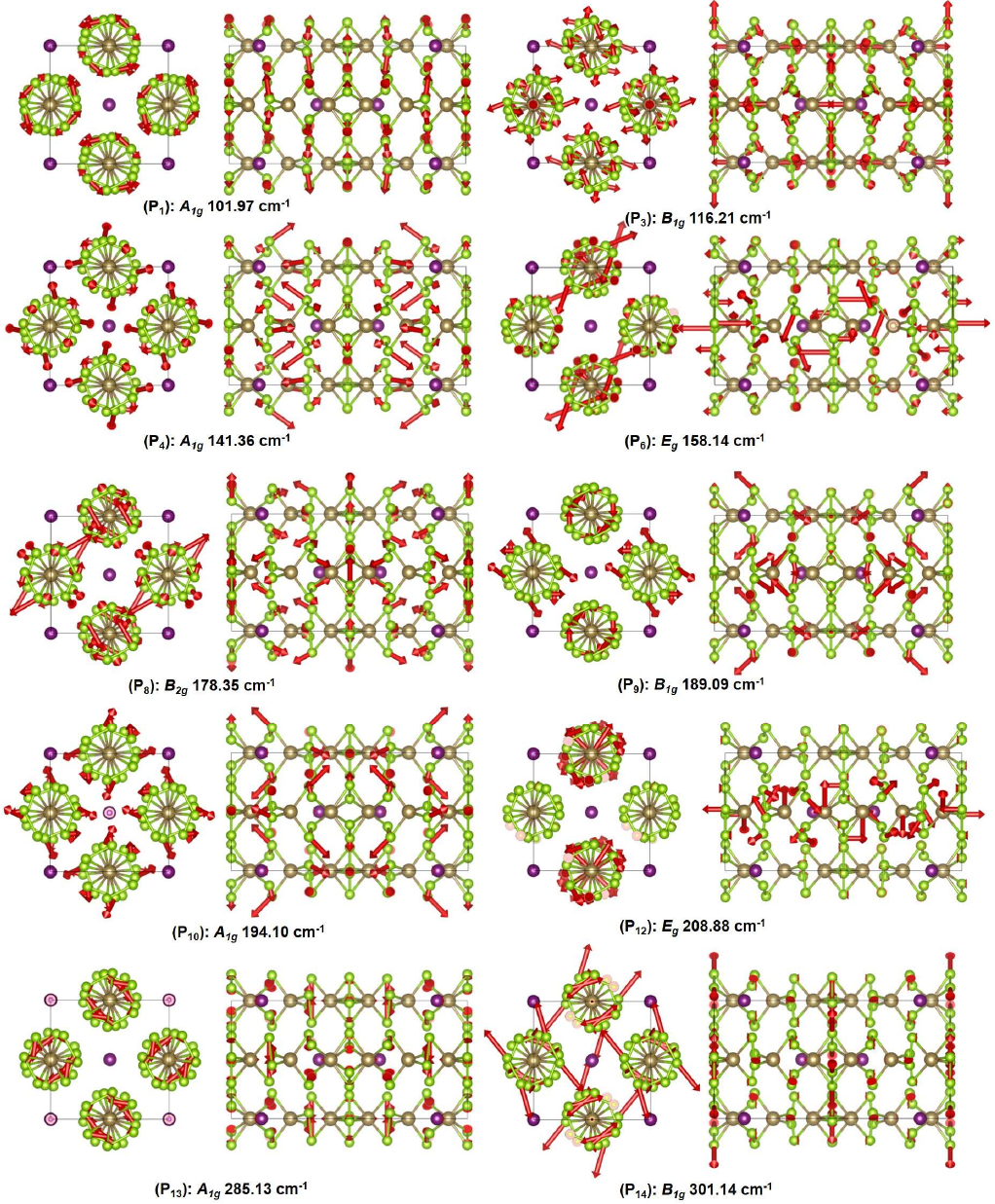}
	\caption {\textbf{Raman Vibration symmetry representation at room temperature}}  
\end{figure*}

\end{document}